\documentclass[12pt]{article}
\usepackage{amsmath}
\usepackage{setspace, lscape}
\RequirePackage{amsthm,amsmath,amsfonts,amssymb,mathtools}
\RequirePackage[authoryear]{natbib}
\RequirePackage[colorlinks,citecolor=blue,urlcolor=blue]{hyperref}
\RequirePackage{graphicx}
\newcommand{\blind}{1}
\usepackage{appendix}
\usepackage{booktabs}

\usepackage{bm}
\usepackage{float}
\usepackage{hhline}
\usepackage{rotating} 
\usepackage{multirow}
\usepackage{adjustbox}
\usepackage[normalem]{ulem}
\usepackage{xr}
\usepackage{subfig}

\newcommand{\true}{{\mbox{\scriptsize tr}}}

\newcommand{\Nor}{\mbox{N}}
\newcommand{\logN}{\mbox{log-N}}
\newcommand{\IG}{\mbox{inv-Ga}}

\newcommand{\Be}{\text{Be}}

\newcommand{\Dir}{\text{Dir}}
\newcommand{\Ga}{\text{Ga}}
\newcommand{\unif}{\text{Unif}}

\newcommand{\bx}{\bm{x}}

\newcommand{\by}{\bm{y}}
\newcommand{\bY}{\bm{Y}}

\newcommand{\bff}{\bm{f}}
\newcommand{\bq}{\bm{q}}

\newcommand{\bmu}{{\bm \mu}}
\newcommand{\bbet}{{\bm \beta}}
\newcommand{\bth}{{\bm \theta}}
\newcommand{\blam}{{\bm \lambda}}
\newcommand{\balpha}{{\bm \alpha}}

\newcommand{\uI}{\mbox{I}}
\newcommand{\Exp}{\mbox{E}}

\newcommand{\Cov}{\mbox{Cov}}

\newcommand{\Prob}{\text{P}}

\newcommand{\iid}{\stackrel{iid}{\sim}}
\newcommand{\indep}{\stackrel{indep}{\sim}}

\usepackage{tabularx}
\usepackage[table]{xcolor}
\usepackage{colortbl}

\begin{document}

\def\spacingset#1{\renewcommand{\baselinestretch}%
{#1}\small\normalsize} \spacingset{1}


\if1\blind
{
  \title{\bf Bayesian Covariate-Varying Interaction Analysis for Multivariate Count Data: Application to Microbiome Studies}
  
  \author{
        Shuangjie Zhang
        \thanks{Address for Correspondence:  105 E 24th St, Austin, TX, 78712. E-mail: shuangjie.zhang@austin.utexas.edu. }\\
        {\small Department of Statistics and Data Sciences, The University of Texas at Austin;}
        \and
        Michael L.\ Patnode \\
        {\small Department of Microbiology and Environmental Toxicology,  University of California Santa Cruz}
        \and
        Juhee Lee \\
        {\small Department of Statistics,  University of California Santa Cruz}
}

  \maketitle
} \fi

\if0\blind
{
  \bigskip
  \bigskip
  \bigskip
  \begin{center}
    {\LARGE\bf Title}
\end{center}
  \medskip
} \fi

\bigskip
\begin{abstract}
\noindent
Understanding covariate-varying interdependencies among features is of great interest in various applications. Motivated by microbiome studies where microbial abundances and interactions vary with environmental factors, we develop a Bayesian covariate-varying factor model. This model flexibly estimates heteroscedasticity in the covariance matrix as a function of covariates. Specifically, our approach employs covariance regression through linear regression on a lower-dimensional factor loading matrix. This formulation, combined with joint sparsity induced by the Dirichlet--Horseshoe prior for the factor loadings, provides robust estimation of covariate-varying covariance in high-dimensional settings. The model simultaneously incorporates a regression structure for the mean abundance and jointly addresses the covariate-varying mean and covariance structure. Furthermore, the model tackles key statistical challenges such as discreteness, over-dispersion, compositionality, and high dimensionality, common in microbiome data analysis, using a flexible nonparametric Bayesian framework. We thoroughly investigate the properties of the model and conduct extensive simulation studies to examine its performance. Real microbiome data examples are provided for illustration.

\end{abstract}

\noindent%
{\it Keywords:} Covariate-varying interdependencies, Microbiome data analysis, Multivariate counts, Bayesian factor model, Covariance regression, Heteroscedasticity. \vfill

\newpage
\spacingset{1.75} 

\section{Introduction}\label{sec:ch4-intro}

Covariance estimation is a core task in multivariate analysis and is essential for characterizing relationships among variables. Covariance matrices for a multivariate random vector play a pivotal role in various statistical methods, including principal component analysis \citep{pearson1901liii}, factor analysis \citep{rummel1988applied}, and canonical-correlation analysis \citep{hotelling1992relations}. Conventional methods for covariance estimation, such as the sample covariance matrix and covariance estimators with structural assumptions \citep{fan2013large}, typically assume that the data are identically and independently distributed (i.i.d.). However, this assumption is frequently violated in practice, as real-world data often exhibit heteroscedasticity, that is, covariances among variables vary with covariates. Ignoring such covariate dependencies can yield misleading conclusions, necessitating the development of methods that account for these dependencies to provide sensible estimates of the interrelationships among variables. The problem is even more challenging for high-dimensional data, whose dimension can be much larger than the sample size.

Covariance regression has long received significant attention due to its ability to incorporate covariate information, improving both the accuracy and interpretability of covariance estimates. \cite{carroll1982robust} first considered a heteroscedastic model in which the variances were given by a parametric function of the mean. Additional approaches include a linear model for the standard deviation \citep{rutemiller1968estimation} and a generalized model with a link function to allow non-negativity of variance \citep{smyth1989generalized} were also developed for univariate cases. When it comes to multivariate heteroscedasticity, \cite{Leonard1992}, \cite{chiu1996matrix}  and \cite{pourahmadi2011covariance} modeled the logarithm of elements of the covariance matrix as a linear function of known matrices to guarantee the positive definiteness of the covariance matrix. However, interpreting covariate effects on the log scale can be challenging, and the number of parameters to estimate grows rapidly in high-dimensional settings. More recently, sparse and low-rank methods for covariate-dependent covariance estimation or its inversion (precision matrix) have been considered to manage high-dimensional data where traditional methods are often inadequate. These approaches leverage structural assumptions such as sparsity and low-rank structure to enhance estimation accuracy and interpretability. \cite{zhang2012moving} modeled Cholesky decomposition of covariance matrix as linear functions of covariates with a moving average construction. \cite{hoff2012covariance} expressed the covariance as a baseline covariance matrix plus a rank-1 positive definite matrix which depends on covariates. They further extended to allow the deviation of each covariate-dependent covariance from the baseline to be any rank. \cite{fox2015bayesian} placed a Gaussian process prior on the latent factor model and induced a flexible Bayesian nonparametric covariance regression model. The predictor-dependent framework was characterized as a combination of Gaussian process random functions of covariates. \cite{ni2019bayesian} proposed a graphical regression method that estimates directed acyclic graphs for the precision matrix in heterogenous data with additional subject-level covariates. \cite{niu2023covariate} further modeled continuously varying undirected graphs with additional assistance from any general covariates for underlying heterogeneous multivariate observations. 

Besides modeling the covariance matrix with covariates, joint modeling for means and covariances allows for the simultaneous exploration of covariate effects on the mean and the covariance of a variable(s) of interest. \cite{pourahmadi1999joint} provided a joint mean-covariance model with applications to longitudinal data. In the context of temporal heteroscedasticity, \cite{fong2006simple} studied multivariate autoregressive conditionally heteroscedastic (ARCH) models in the financial data. \cite{niu2019joint} extended their model in \cite{hoff2012covariance} to a joint mean and covariance model, studying the covariate effects on both mean and covariance in the application of multiple health outcome measures. \cite{moran2021bayesian} used a parametric covariance regression model to analyze verbal autopsy data. It was designed specifically for cause of death denoted covariance. However, the above approaches are built for continuous data, and they can be inappropriate for analyzing multivariate count data.

With the advent of high-throughput sequencing (HTS) technologies, multivariate count tables arise in various biological applications for statistical analyses.  Especially in microbiome studies, 16S ribosomal RNA (16S rRNA) sequencing uses similarity-based clustering algorithms to group 16S rRNA sequences into Operational Taxonomic Units (OTUs), producing multivariate count tables for downstream analysis. An OTU represents a group of organisms classified together based on genetic sequence similarity. Analyzing OTU tables and detecting the structure of microbial interactions is essential for more accurately characterizing microbial communities. 
Popular methods in microbiome studies such as SparCC \citep{friedman2012inferring}, CCLasso \citep{fang2015cclasso} and SPIEC-EASI \citep{kurtz2015sparse} adopt log-transformed counts or log-transformed ratios for analysis of interactions. Specifically, SparCC \citep{friedman2012inferring} adds pseudo counts and then divides the raw counts by the sample's total counts for normalization. It models log-transformed ratios of these normalized counts to infer correlations between OTUs through sparse networks. Similarly, CCLasso in \cite{fang2015cclasso} uses $\ell_1$ penalty to estimate the correlation network of log-transformed counts. SPIEC-EASI \citep{kurtz2015sparse} uses graphical lasso \citep{friedman2008sparse}, a popular penalized method outputting the association of undirected graphs, to obtain a robust precision matrix estimate. The raw OTU counts are also first centered by log-ratio (clr) transformation. See  REBECCA \citep{ban2015investigating}, COAT \citep{cao2019large} and MOFA \citep{argelaguet2018multi} for more. However, most existing methods merely center the data at the sample mean and assume a constant covariance structure across all covariates.

To circumvent these challenges and model the influence of covariates on microbial interactions, we propose a Bayesian covariate-varying sparse factor model employing a rounded-kernel formulation. The model assesses interrelationships between OTUs varying as a function of covariates. Furthermore, it simultaneously performs model-based normalization and flexibly accommodates large variability in count data. Specifically, we use nonparametric mixtures of rounded multivariate log-normal kernels to introduce latent continuous random variables. The covariance matrix of the kernels is then allowed to vary with covariates, characterizing the covariate-varying interrelationships among OTUs. We adopt the low-rank structure factor model and place a Dirichlet–Horseshoe (Dir-HS) prior \citep{zhang2023} on the factor loading matrix to effectively learn a high-dimensional covariance structure despite a limited sample size. 
In addition, we use a Dirichlet process (DP) prior on microbial abundances to obtain a flexible joint distribution of count vectors. This nonparametric approach used to model the mean counts flexibly handles excess zeros and overdispersion, common in microbiome data. We also relate covariates to the mean to detect different OTU abundances under covariates. 

In the rest of the paper, we describe the model and its applications. Sections~\ref{sec:ch4-model} and \ref{sec:ch4-mcmc} present the covariate-dependent rounded multivariate log-normal kernel model, its prior specification, and posterior computation. Section~\ref{sec:ch4-sim} shows the results of simulation studies evaluating the performance of our method. Section~\ref{sec:ch4-data} presents results from applying the model to a real dataset, and Section~\ref{sec:ch4-con} concludes with a discussion of the findings and directions for future research.

\section{Model Specification}\label{sec:ch4-model}

In this section, we first construct a Bayesian sparse factor model that allows factor loadings to vary with covariates, enabling the estimation of covariate-varying covariance matrices. We then develop a rounded kernel model with a mean regression function to accommodate the discreteness of count data. This model employs a Bayesian nonparametric approach, providing a flexible joint distribution for multivariate count responses.

\subsection{Sparse Covariate-dependent Factor Model}\label{sec:ch4-1}

Let $\tilde{\bY}^\star = (\tilde{Y}^\star_{1}, \ldots, \tilde{Y}^\star_{J}) \in \mathbb{R}^J$ be a $J$-dimensional normal random vector,
\begin{equation}
\tilde{\bY}^\star \mid \bmu(\bx), \Sigma(\bx) \;\sim \; \mathrm{N}_J\!\big(\bmu(\bx),\, \Sigma(\bx)\big),
\label{ch4-eq:y_star_tilde}
\end{equation}
where $\bx = (x_{1}, \ldots, x_{P})$ is a $P$-dimensional vector of covariates for the sample, with $x_{1} = 1$ representing the intercept. 
We first construct a probability model for the covariance matrix $\Sigma(\bx)$ as a function of $\bx$, which is the main parameter of interest. 
The covariates included in $\bmu(\bx)$ may differ from those in $\Sigma(\bx)$ depending on the context of the problem.  A model for $\bmu(\bx)$ will be discussed later in \S~\ref{sec:ch4-2}.
To overcome the difficulty posed by high dimensionality, particularly when the sample size is much smaller than the number of features, i.e., $N \ll J$ for each value of $\bx$, we extend the spiked covariance structure assumption \citep{johnstone2001distribution}. Specifically, we decompose $\Sigma(\bx)$ into a low-rank matrix and a diagonal matrix, and construct a factor model whose loadings vary with $\bx$:
\begin{equation} 
\Sigma(\bx) = \Lambda(\bx)\Lambda^\prime(\bx) + \sigma^2 \uI_J,
\label{eq:ch4-Sigma-decomp}
\end{equation}
where $\Lambda(\bx) = [\lambda_{jk}(\bx)]$, $j = 1,\ldots,J$ and $k = 1,\ldots,K$, is a $J \times K$ covariate-dependent factor loading matrix. Here, $K$ is the dimension of the subspace that is assumed to capture the statistical variability, and typically $K \ll J$. Following \cite{bhattacharya2011sparse} and \cite{xie2018bayesian}, we do not impose any constraints on $\Lambda(\bx)$, such as column orthogonality, nor do we seek to interpret latent factors, as our primary focus is on inference for $\Sigma(\bx)$.

We further express $\lambda_{jk}(\bx)$ as
\begin{equation} 
\lambda_{jk}(\bx) = q_{jk}\,\bff_k^\prime \bx.
\label{eq:ch4-lambda-decomp}
\end{equation}
Here, $\bff_k$ is a $P$-dimensional coefficient vector that quantifies covariate effects for factor $k$. While the global effect $\bff_k^\prime \bx$ is shared across features, $q_{jk}$ adjusts its contribution for feature $j$. When the local effect $q_{jk}$ is close to 0, the corresponding $\lambda_{jk}(\bx)$ becomes small. On the other hand, if $f_{kp} \approx 0$, then $\lambda_{jk}(\bx)$ does not vary with $x_p$ for all $j$. If $f_{kp}$ is small for all $k$ and $p$, then $\Sigma(\bx)$ does not change much with $\bx$.

Using \eqref{eq:ch4-Sigma-decomp} and \eqref{eq:ch4-lambda-decomp}, we can rewrite
\begin{equation}
\Sigma(\bx)
  = \sum_{k=1}^K (\bq_k \bff_k^\prime \bx)(\bx^\prime \bff_k \bq_k^\prime) 
    + \sigma^2 \uI_J,
\label{eq:cov-reg}
\end{equation}
where $\bq_k = (q_{1k}, \ldots, q_{Jk})^\prime$. The covariance between features $j$ and $j^\prime$ is a sum of $K$ quadratic functions of $\bx$:
\begin{equation} 
\Sigma_{jj^\prime}(\bx)
  = \sum_{k=1}^K q_{jk} q_{j^\prime k} \big(\bff_k^\prime \bx\big)^2
    + \sigma^2\, 1(j = j^\prime),
\label{eq:ch4-lambda-cov}
\end{equation}
which can flexibly capture various shapes over a reasonable range of $\bx$. Similar to \cite{hoff2012covariance} and \cite{niu2019joint}, including the intercept $1$ in $\bx$ alleviates the constraint that the minimum covariance always occurs at $x_{p} = 0$ for $p > 1$.
The structure in \eqref{eq:cov-reg} has $JK + KP + 1$ unknown parameters and significantly reduces the number of parameters to estimate compared to estimating a full $J \times J$ covariance matrix $\Sigma$ for each $\bx$. This reduction is crucial in high-dimensional settings.

In addition to the spiked covariance structural assumption, we further assume joint sparsity on $\Sigma(\bx)$ by employing the Dirichlet--Horseshoe (Dir-HS) prior of \cite{zhang2023} for $\bq_k$, for $k = 1,\ldots,K$:
\begin{eqnarray}
\begin{aligned}
\tau_{k} \mid a_\tau, b_\tau &\iid \Ga(a_\tau, b_\tau/J), \\
{\bm \phi}_{k} = (\phi_{1k}, \ldots, \phi_{Jk}) \mid a_{\phi} &\iid \Dir(a_{\phi}, \ldots , a_{\phi}), \\
\zeta_{jk} &\iid \mbox{C}^{+}(0,1), \qquad j=1,\ldots,J,\\
q_{jk} \mid \phi_{jk}, \tau_{k}, \zeta_{jk} &\indep \Nor(0, \zeta_{jk}^2\, \phi_{jk}\tau_{k}), \qquad j=1,\ldots,J.
\end{aligned}
\label{eq:ch4-prior-lam}
\end{eqnarray}
Here, $\mbox{C}^{+}(0,1)$ denotes the half-Cauchy distribution on $\mathbb{R}^+$ with location $0$ and scale $1$, and $\Ga(a,b)$ represents the gamma distribution with mean $a/b$.  
Under the model in \eqref{eq:cov-reg} with the prior in \eqref{eq:ch4-prior-lam}, the vector $\bm{\phi}_k$ implicitly induces feature selection for interactions for latent factor $k$, resulting in the shrinkage of $q_{jk}$ corresponding to irrelevant features toward zero. This, in turn, shrinks $\lambda_{jk}(\bx)$, and those irrelevant features exhibit a small covariance regardless of $\bx$. In contrast, $\tau_k$ controls global shrinkage for each factor and effectively truncates the number of active latent factors. This joint sparsity assumption facilitates more reliable estimation of the covariance structure with limited sample sizes and yields favorable theoretical properties \citep{cai2015optimal, xie2018bayesian}. \cite{zhang2023} further discussed the theoretical properties of the Dir-HS distribution, showing that it achieves heavy tails together with a sharp peak at zero, thereby efficiently inducing joint sparsity and enabling robust covariance estimation in high-dimensional settings.  

Compared to the covariance-regression models discussed earlier \citep{fox2015bayesian, moran2021bayesian, hoff2012covariance}, our model is more parsimonious due to \eqref{eq:ch4-lambda-decomp}, while the form in \eqref{eq:cov-reg} remains flexible enough to capture nontraditional patterns. Moreover, the structural assumption with joint sparsity yields more efficient and robust estimation in high dimensions. We illustrate this in \S~\ref{sec:ch4-sim2} with simulation studies in which the true underlying covariance matrix is an arbitrary function of a covariate.  

To complete the prior specification for $\Sigma(\bx)$, we assume a conditionally conjugate prior on $\sigma^2$,  
$
\sigma^2 \sim \IG(a_\sigma, b_\sigma),
$
with fixed hyperparameters $a_\sigma$ and $b_\sigma$, and a standard normal prior on $f_{kp}$,  
$
f_{kp} \iid \Nor(0,1).
$

\subsection{A Flexible Model for Multivariate Count Responses}\label{sec:ch4-2}

Next, we use the latent factor models with covariate-varying factor loadings in \S~\ref{sec:ch4-1} as a building block to construct a Bayesian nonparametric mixture model with a rounded kernel, yielding a flexible multivariate count distribution whose mean and covariance vary with $\bx$. 

Suppose that we have $N$ samples, for which we observe $J$-dimensional random count vectors $\bY_{i}=(Y_{i1}, \ldots, Y_{iJ})$, with $Y_{ij} \in \mathbb{Z}_{\ge 0}$, $i=1, \ldots, N$, corresponding to features with covariates $\bx_i$. We introduce a latent multivariate log-normal vector $\bY^\star_i = \exp(\tilde{\bY}^\star_i) \in \mathbb{R}_+^J$ and assume that the $\tilde{\bY}^\star_i$'s are independent samples from the model in \eqref{ch4-eq:y_star_tilde}. We then construct a joint distribution for the count vector $\bY_i$ using a rounded kernel \citep{canale2011bayesian}:
\begin{eqnarray}
\begin{aligned}
\Prob(\bY_i = \by \mid \bmu(\bx_i), \Sigma(\bx_i)) &=& \int_{A(\by)} f_{\by^\star}(\by^\star \mid  \bmu(\bx_i), \Sigma(\bx_i)) d\by^\star,
\end{aligned}
\label{ch4-eq:dist-y}
\end{eqnarray} 
where the integration region is $A(\by)=\{\by^\star \mid y_{1} \leq y^\star_{1} < y_{1}+1, \ldots, y_{J} \leq y^\star_{J} < y_{J}+1\}$, and $f_{\by^\star}(\cdot)$ is the probability density function of a $J$-dimensional log-normal distribution with parameters $\bmu(\bx_i)$ and $\Sigma(\bx_i)$. The multivariate log-normal density is zero for vectors with negative entries, and the kernel defines a valid multivariate count distribution for $\bY_i$. The quantity $\exp(\mu_{j}(\bx_i))$ is the median of $Y^\star_{ij}$ and provides an estimate of the abundance of feature $j$ in sample $i$. We adopt the model in \S~\ref{sec:ch4-1} for $\Sigma(\bx_i)$. The mean and covariance of $Y^\star_{ij}$ are
\[
\Exp(Y^\star_{ij}) = \exp\Big(\mu_{j}(\bx_i) + \frac{1}{2}\Sigma_{jj}(\bx_i)\Big), \quad
\Cov(Y^\star_{ij}, Y^\star_{ij^\prime}) = \Exp(Y^\star_{ij}) \Exp(Y^\star_{ij^\prime}) \Big\{\exp(\Sigma_{jj^\prime}(\bx_i)) -1\Big\}.
\]
When $\Sigma_{jj^\prime}(\bx_i) = 0$, it implies there are no microbial interactions between features $j$ and $j^\prime$. For the count distribution $\bY_i$, the mean and covariance are finite and can be computed through the probability mass function defined in \eqref{ch4-eq:dist-y}. We illustrate the distributions of $\by$ obtained under the rounded kernel model with specific examples in Supp.~\S2.

We relate $\bmu(\bx_i)$ to covariates $\bx_i$ through the regression
\begin{equation}
\mu_{j}(\bx_i) = \mu_{ij} = r_i + \alpha_{j} + \bbet_j^\prime \tilde{\bx}_i, 
\label{eq:ch4-mu} 
\end{equation}
where $\tilde{\bx}_i$ denotes $\bx_i$ without the intercept $x_{i1}$. The term $r_i$ is the sample (library) size factor, which normalizes counts across samples, and $\alpha_{j}$ represents the normalized baseline abundance of feature $j$. The regression coefficients $\beta_{jp}$ quantify the change in the abundance of feature $j$ from its baseline $r_i + \alpha_{j}$ due to covariate $x_{ip}$. We consider a conditionally conjugate prior distribution for $\beta_{jp}$ and assume $\beta_{jp} \iid \Nor(0, u^2_\beta)$ with fixed $u^2_\beta$.

We take a Bayesian nonparametric approach and construct a flexible prior model for $\alpha_j$ to account for the large variability in abundance among features. This flexible modeling of baseline abundances may further enhance the estimation of $\Sigma(\bx_i)$ and $\bbet_j$. While $\bbet_j$ and $r_i + \alpha_j$ are identifiable, the individual parameters $r_i$ and $\alpha_j$ in \eqref{eq:ch4-mu} are not identifiable due to the multiplicative structure, $\Exp(y^\star_{ij} \mid r_{i}, \alpha_{j}) \propto \exp(r_{i} + \alpha_{j})$. However, since the primary inferential goal is to estimate $\beta_{jp}$, this non-identifiability does not impact the inference of interest. To mitigate potential computational issues, we impose a mean-constrained Dirichlet process prior for $\alpha_j$ as follows:
\begin{eqnarray}
\begin{aligned}
\alpha_j \mid G &\iid G = \sum_{l=1}^{\infty} \psi_{l}^\alpha 
\left\{ \omega_{l}^\alpha \delta_{\xi_{l}^\alpha} + (1-\omega_{l}^\alpha) \delta_{\frac{\nu^\alpha - \omega_{l}^\alpha \xi_{l}^\alpha}{1-\omega_{l}^\alpha}} \right\}, \quad j=1, \ldots, J,
\label{eq:ch4-prior-balp} 
\end{aligned}
\end{eqnarray}
where $\delta_\xi$ is a point mass centered at $\xi$. We let $\xi_{l}^\alpha \mid \nu^\alpha, u^2_\alpha \iid \Nor(\nu^\alpha, u^2_\alpha)$, $l=1, 2, \ldots$, with fixed $\nu^\alpha$ and $u^2_\alpha$.  
The outer mixture weights $\psi_{l}^\alpha$ in \eqref{eq:ch4-prior-balp} are constructed using a stick-breaking process \citep{sethuraman1994constructive}: let $\psi_{1}^\alpha = V^\alpha_{1}$ and $\psi_{l}^\alpha = V^\alpha_{l} \prod_{l^\prime=1}^{l-1} (1-V^\alpha_{l^\prime})$ for $l > 1$, with $V^\alpha_{l} \mid c^\alpha \iid \Be(1, c^\alpha)$, where the concentration parameter $c^\alpha$ is fixed. 
We assume the inner mixture weights satisfy $\omega^\alpha_{l} \mid a^{\alpha}_\omega, b^{\alpha}_\omega \iid \Be(a^{\alpha}_\omega, b^{\alpha}_\omega)$, with fixed $a^{\alpha}_\omega$ and $b^{\alpha}_\omega$. Under \eqref{eq:ch4-prior-balp}, both the prior and posterior means of $\alpha_j$ are fixed at $\nu^\alpha$. A similar constraint is imposed on the prior for $r_i$ to achieve soft identifiability. \cite{shulerbayesian} and \cite{zhang2023} showed that overall means can be well estimated under such mean-constrained priors.  

Under the model in \eqref{ch4-eq:y_star_tilde} and \eqref{eq:ch4-prior-balp}, we obtain a Dirichlet process mixture model for $\bY^\star_i$:
\begin{eqnarray}
\bY^\star_i \mid \bmu(\tilde\bx_i), \Sigma(\bx_i) &\indep& \int \mbox{log-N}_J \Big(\by^\star \mid r_i \bm{1}_J + \balpha + \bbet \tilde\bx_i^\prime, \Sigma(\bx_i) \Big) \prod_j dG(\alpha_j),
\label{ch4-eq:y_star-dpm}
\end{eqnarray}
where $\balpha = (\alpha_1, \ldots, \alpha_J)^\prime$ and $\bbet$ is a $J \times P$ coefficient matrix with $\bbet_j$ in the rows. With random mixture weights $\omega_l^\alpha$ and $\psi_l^\alpha$ and random locations $\xi_l^\alpha$ in $G(\alpha)$, the mixture model in \eqref{ch4-eq:y_star-dpm} can flexibly capture various shapes of the distribution and accommodate variability in the count data. We also consider an extension of the model in \eqref{eq:ch4-mu}–\eqref{ch4-eq:y_star-dpm} to account for inter-subject heterogeneity, illustrated in detail in Simulations 2 and 3.  

Similar to \eqref{eq:ch4-prior-balp}, we consider a flexible infinite mixture model for $r_i$:
\begin{eqnarray}
r_i \mid \psi_l^r, \omega_l^r &\iid& \sum_{l=1}^\infty \psi_l^r \left\{ 
\omega_l^r \Nor(\xi_l^r, u_r^2) + (1-\omega_l^r) \Nor\left(\frac{\nu^r - \omega_l^r \xi_l^r}{1-\omega_l^r}, u_r^2\right) 
\right\},
\label{eq:ch4-prior-ri}
\end{eqnarray}
where $\nu^r$ and $u_r^2$ are fixed. The prior and posterior expectations of $r_i$ are fixed at $\nu^r$ in \eqref{eq:ch4-prior-ri}. We jointly specify $\nu^\alpha$ and $\nu^r$ using observed counts. For example, we set $\nu^r$ to the average logarithm of the total count, $\nu^r = \frac{1}{N} \sum_{i=1}^{N} \log\Big(\sum_{j=1}^{J} y_{ij}\Big)$, and $\nu^\alpha = \frac{1}{NJ} \sum_{i=1}^{N} \sum_{j=1}^{J} \big\{\log(y_{ij}+0.01) - \nu^r \big\}$.  
We assume similar priors for $\psi_l^r$, $\omega_l^r$, and $\xi_l^r$: $\xi_l^r \mid \nu^r, u_{\xi^r}^2 \iid \Nor(\nu^r, u_{\xi^r}^2)$, $\omega_l^r \mid a_\omega^r, b_\omega^r \iid \Be(a_\omega^r, b_\omega^r)$, and $\psi_1^r = V_1^r$, $\psi_l^r = V_l^r \prod_{\ell'=1}^{l-1} (1-V_{\ell'}^r)$ for $l>1$, with $V_l^r \mid c^r \iid \Be(1, c^r)$. Here, $u_{\xi^r}^2$, $a_\omega^r$, $b_\omega^r$, and $c^r$ are fixed.

\section{Prior Calibration and Posterior Computation}\label{sec:ch4-mcmc}

The prior of $\Sigma(\bx)$ in \eqref{eq:ch4-prior-lam} requires the specification of fixed hyperparameters $K$, $a_\phi$, $a_\tau$, and $b_\tau$. Selecting an appropriate dimension $K$ for the latent space can be challenging. The value of $K$ determines the number of parameters, and a model with a random $K$ would require more complex algorithms, such as a reversible jump Markov Chain Monte Carlo (RJMCMC) method \citep{green2009reversible}, for posterior simulation. Instead, we set $K$ to a reasonably large value for computational convenience. For example, we empirically determine $K$ by performing principal component analysis (PCA) on the sample covariance matrix of log-transformed normalized counts. We then choose $K$ such that the $K$ largest eigenvalues explain 95\% of the total variance. With a sufficiently large $K$, the model can allow some $\tau_k$ to be small for redundant latent factors. If desired, a geometric or truncated Poisson prior can be placed on $K$ to achieve an optimal posterior contraction rate \citep{pati2014posterior}.  
For the remaining hyperparameters $a_\phi$, $a_\tau$, and $b_\tau$, we follow the configuration in \cite{zhang2023}, setting $a_\phi = 1/(0.2 \times J)$, $a_\tau = 0.1$, and $b_\tau = 1/J$. From simulation studies, we observed that too small a value of $a_\phi$ tends to overly shrink $q_{jk}$ toward zero, resulting in poor estimates of $\Sigma(\bx)$. We also examined the sensitivity to these hyperparameter specifications and found that the model's performance remains robust within a reasonable range of values.  

Let $\bth = \{q_{jk}, \phi_{jk}, \tau_k, \zeta_{jk}, f_{kp}, \sigma^2, \alpha_j, \omega_l^\alpha, V_l^\alpha, \xi_{jl}^\alpha, r_i, \omega_l^r, V_l^r, \xi_l^r, \beta_{jp}\}$ denote the full set of random parameters. We use Markov Chain Monte Carlo (MCMC) to sample $\bth$ from their posterior distribution. To facilitate computation, we introduce a latent normal vector $\bm{\eta}_i \iid \Nor_K(0, \uI_K)$, so that
$
Y^\star_{ij} \mid \mu_{ij}(\bx_i), \blam_j(\bx_i), \bm{\eta}_i, \sigma^2 \indep \mbox{log-N}\Big(\mu_{j}(\bx_i) + \blam_j^\prime(\bx_i)\bm{\eta}_i, \sigma^2\Big),
$
as independent log-normal variables, which significantly improves computational efficiency.  
The joint posterior distribution of the augmented model is
\begin{eqnarray}
p(\bth, \bY^\star, \bm{\eta} \mid \by, \bx) \propto \prod_{i=1}^N \prod_{j=1}^{J} p\big( y_{ij} \leq Y^\star_{ij} < y_{ij}+1 \mid \bm{\eta}_i, \bth \big) \prod_{i=1}^N p(\bm{\eta}_i \mid \bth) p(\bth).
\label{eq:ch4-model-augmented}
\end{eqnarray} 
We use the blocked Gibbs sampling algorithm \citep{ishwaran2001gibbs} with a finite-dimensional truncation of the stick-breaking processes in \eqref{eq:ch4-prior-balp} and \eqref{eq:ch4-prior-ri}. The truncation levels $L^r$ and $L^\alpha$ are set to sufficiently large values. Given the latent variables, all parameters except $\bm{\phi}_k$ can be updated via Gibbs steps. Although $f_{kp}$ has a conjugate full conditional distribution, we found that mixing over $f_{kp}$ can be poor, so we use an adaptive Metropolis-Hastings algorithm \citep{haario2001adaptive} for efficient updates of $\bm{\phi}_k$ and $f_{kp}$. Details of the MCMC algorithm are provided in Supp.~\S~1. The code is available at \url{https://github.com/shuang-jie/BCAIA}.

\section{Simulation Studies}\label{sec:ch4-sim}

\subsection{Simulation 1}\label{sec:ch4-sim1}

\begin{figure}[!t]
  \begin{center}
\begin{tabular}{ccc}   \includegraphics[width=.33\textwidth]{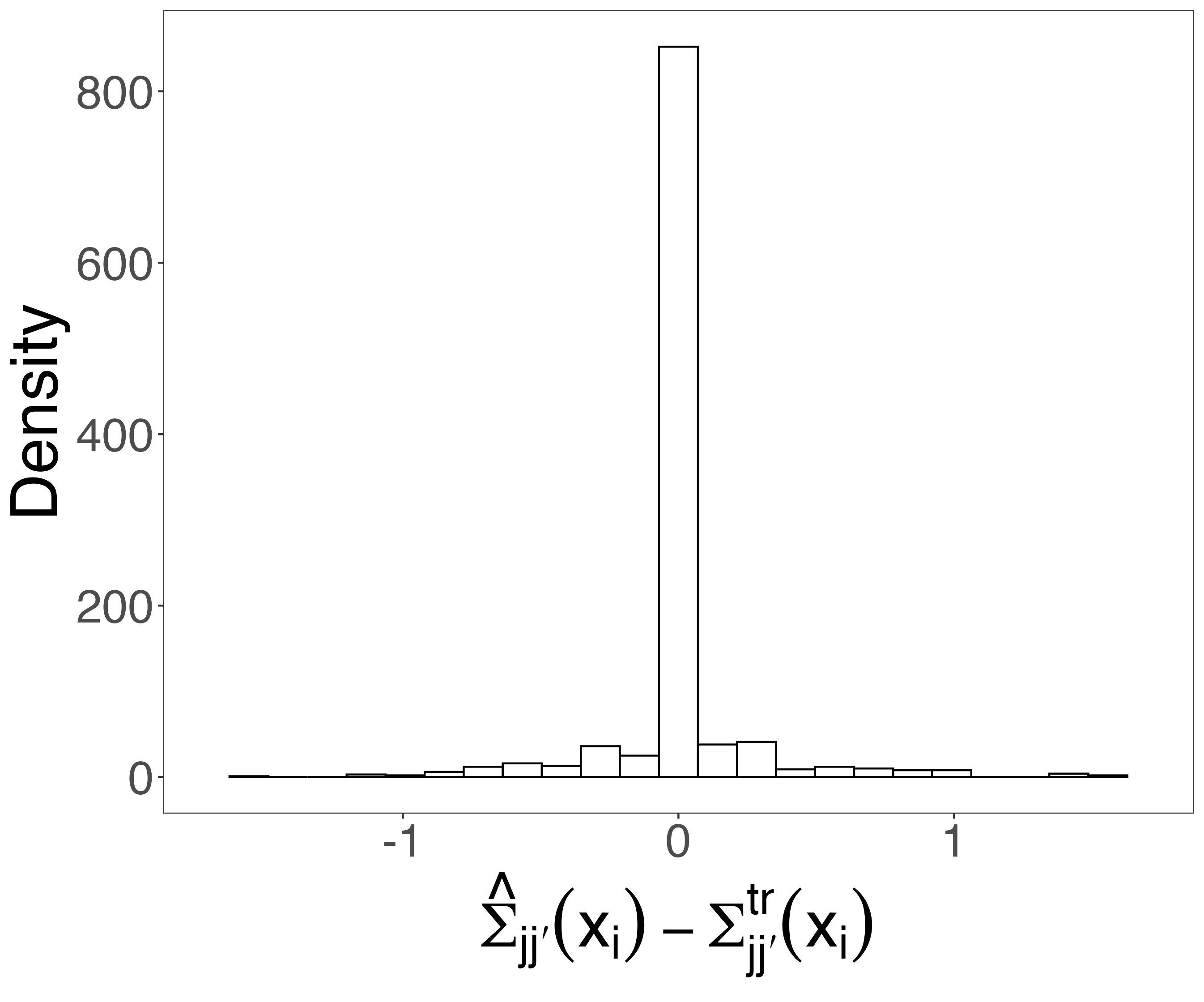} & 
   \includegraphics[width=.32\textwidth]{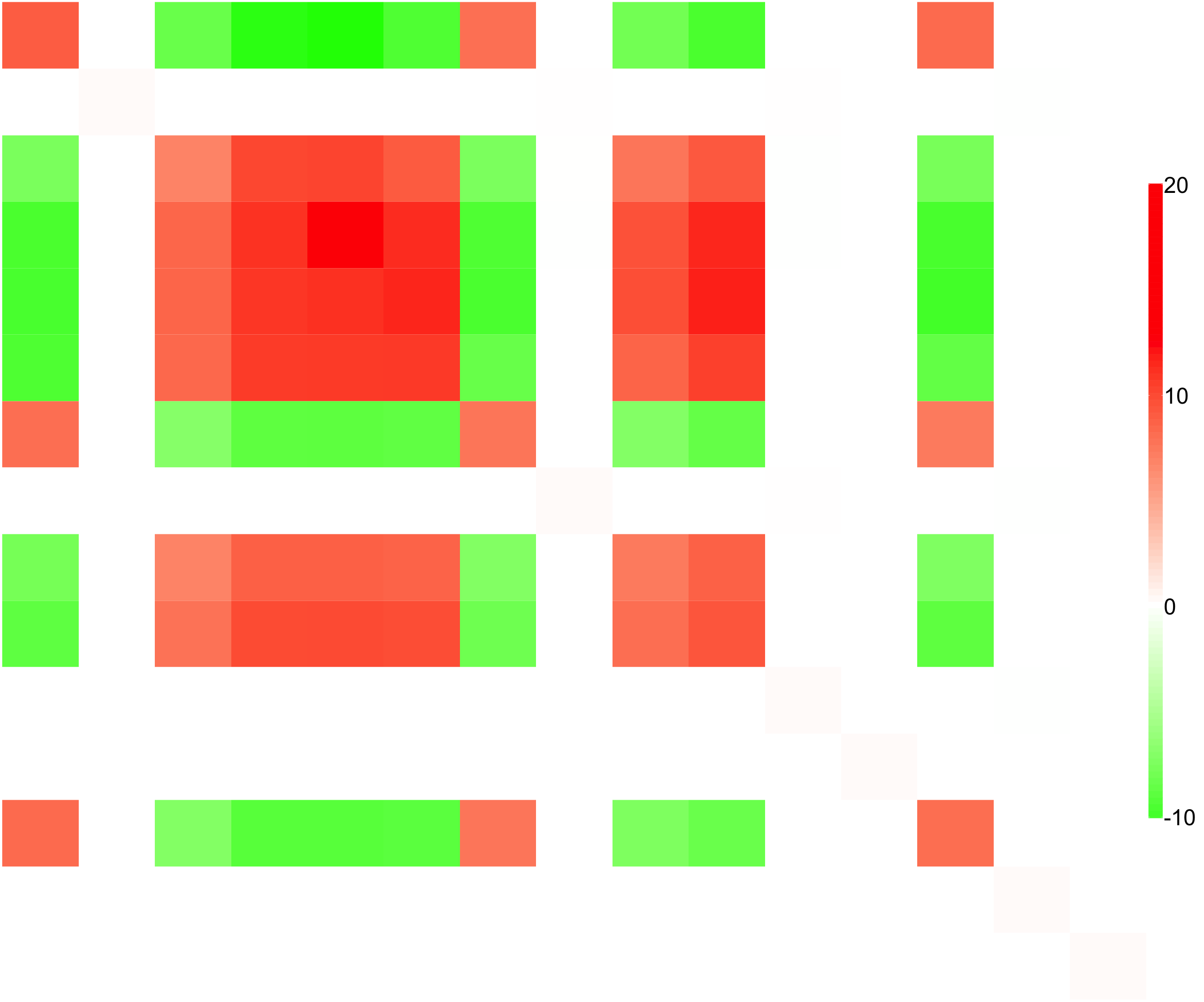} &
   \includegraphics[width=.32\textwidth]{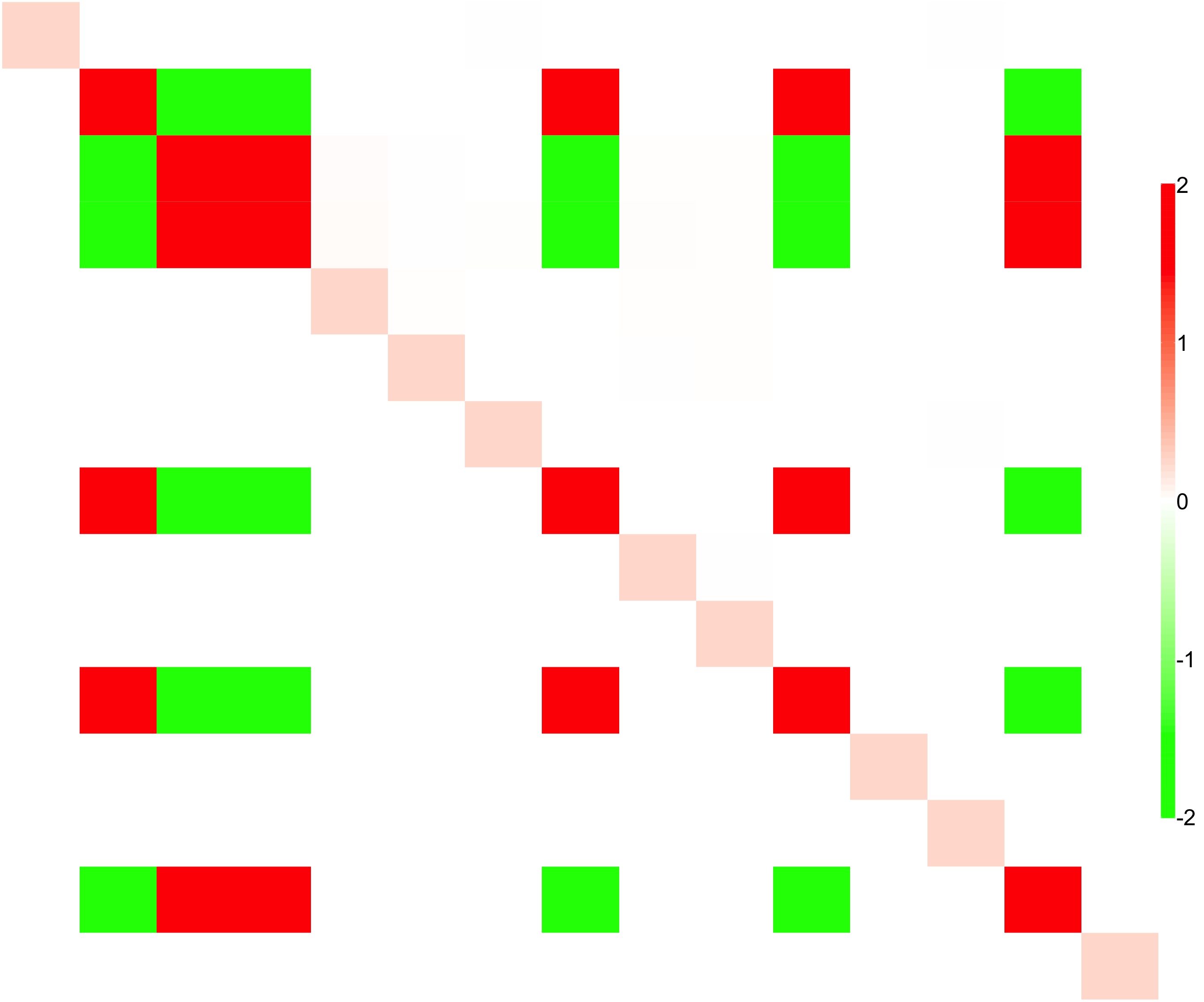} \\
   (a) $\hat\Sigma_{jj^\prime}(\bx)-\Sigma^{\true}_{jj^\prime}(\bx)$ & (b) $\hat{\Sigma}(\bx)$ vs ${\Sigma}^\true(\bx)$ & (c) $\hat{\Sigma}(\bx^\prime)$ vs ${\Sigma}^\true(\bx^\prime)$
  \end{tabular}
 \end{center}
 \vspace{-0.15in}
 \caption{[Simulation 1] Panel (a) has a histogram of differences between $\hat\Sigma_{jj^\prime}(\bx_i)$ and $\Sigma^{\true}_{jj^\prime}(\bx_i)$ under six levels, $j \leq j^\prime$. Panels (b) and (c) compare $\hat{\Sigma}^\true(\bx)$ (lower triangular) to its posterior median estimates $\hat{\Sigma}(\bx)$ (upper triangular) for two arbitrarily selected conditions, $\bx=(1, 0, 0, 1)^\prime$ and $\bx^\prime=(1, 1, 0, 0)^\prime$. } 
\label{fig:ch4-sim2-1-F1}
\end{figure}

Simulation 1 was designed to mirror the structure of the mice gut microbiome dataset in \S~\ref{sec:ch4-data}. The dataset includes two categorical variables, one with two levels and one with three, yielding six experimental conditions. Using indicator variables for the main effects, we constructed $\bx$ with $P=4$ (including the intercept), assumed $J=15$ OTUs, and generated five samples per condition, for a total of $N=30$ samples.
For $\Sigma^\true(\bx)$, we set $K^\true = 2$. Each $q^\true_{jk}$ was set to 0 with probability 0.5; otherwise, it was sampled from $\Nor(0,1)$ and shifted away from zero by 1. We generated $f^\true_{kp} \iid \unif(-1,1)$ and modified some entries so that, under certain conditions, feature interactions depend on only one factor. Specifically, we set $f_{11}=-f_{13}$ so that factor 1 has no contribution to feature interactions for $\bx=(1,0,1,0)$, as $\bm f_1^\prime \bx=0$, and $f_{21}=-f_{22}$ so that factor 2 has no contribution for $\bx=(1,1,0,0)$, as $\bm f_2^\prime \bx=0$. The covariance is then $\Sigma^\true(\bx) = \Lambda^\true(\bx)\Lambda^{\true,\prime}(\bx) + \sigma^{\true,2}\uI_J$, with $\lambda^\true_{jk}(\bx) = q^\true_{jk} \bm f_k^{\true,\prime} \bx$ and $\sigma^{\true,2} = 0.5^2$. True covariance matrices for two arbitrarily selected conditions, $\bx=(1,1,0,0)$ and $\bx=(1,1,0,1)$, are shown in the lower triangles of Fig.~\ref{fig:ch4-sim2-1-F1}(b)-(c).
For the mean abundance, $r_i^\true \iid \unif(0,2)$ and $\alpha_j^\true \iid 0.3\Nor(-1,1^2) + 0.7\Nor(5,0.5^2)$. The coefficients $\beta^\true_{jp}$ were generated similarly to $q^\true_{jk}$: set to 0 with probability 0.5, otherwise sampled from $\Nor(0,1)$ and shifted by 1. Finally, $\bY_i^{\star,\true} \iid \logN_J(\bmu^\true(\bx_i), \Sigma^\true(\bx_i))$, with $\mu_{ij}^\true = r_i^\true + \alpha_j^\true + \bbet_j^\true \tilde{\bx}_i$, and the observed counts are $\bY_i = \lfloor \bY_i^{\star,\true} \rfloor$.

We specified the values of the fixed hyperparameters as discussed in §~\ref{sec:ch4-mcmc}, $K=8$, $c^r=c^\alpha=3$, $L^r = 30, L^\alpha = 35$, $a_\sigma=b_\sigma=3$, $a^r_\omega=b^r_\omega=a^\alpha_\omega=b^\alpha_\omega=5$ to fit the model for the simulated dataset. We ran MCMC for 160,000 iterations and discarded the first half for burn-in. The computation took 13 minutes on an Apple M1 chip laptop.

\begin{figure}[!t]
  \begin{center}
\begin{tabular}{ccc}
   \includegraphics[width=.33\textwidth]{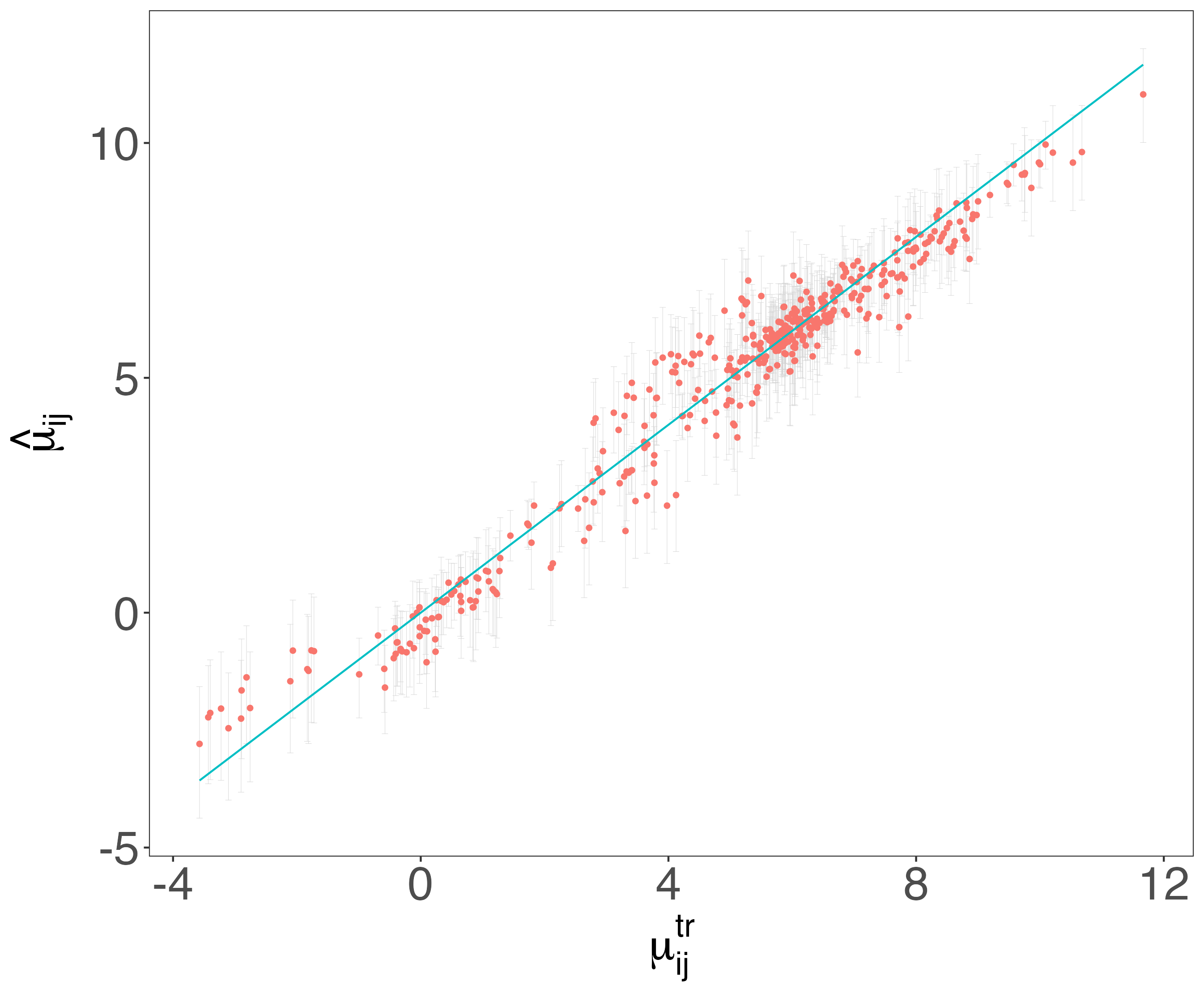} & 
   \includegraphics[width=.33\textwidth]{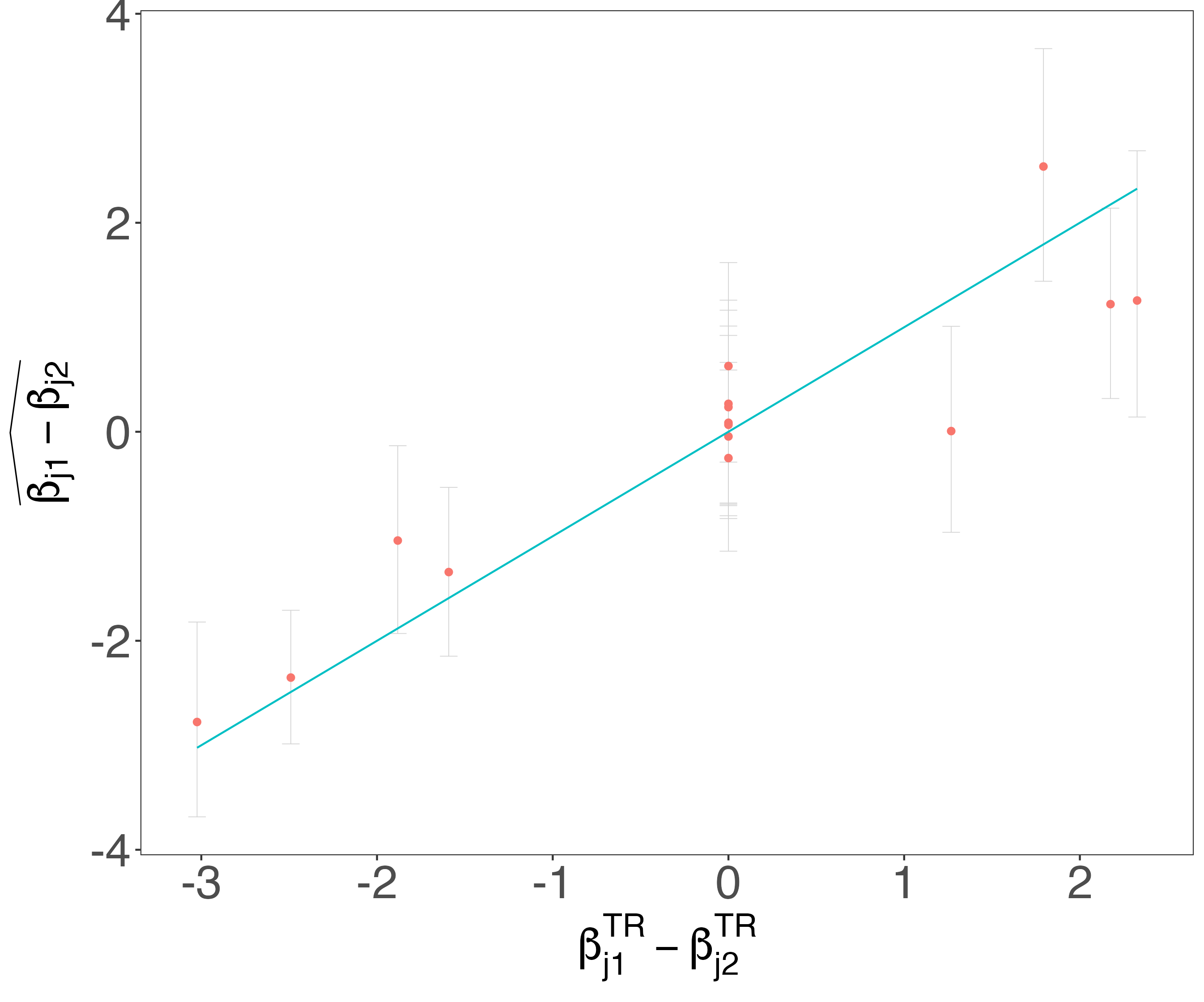} & \includegraphics[width=.33\textwidth]{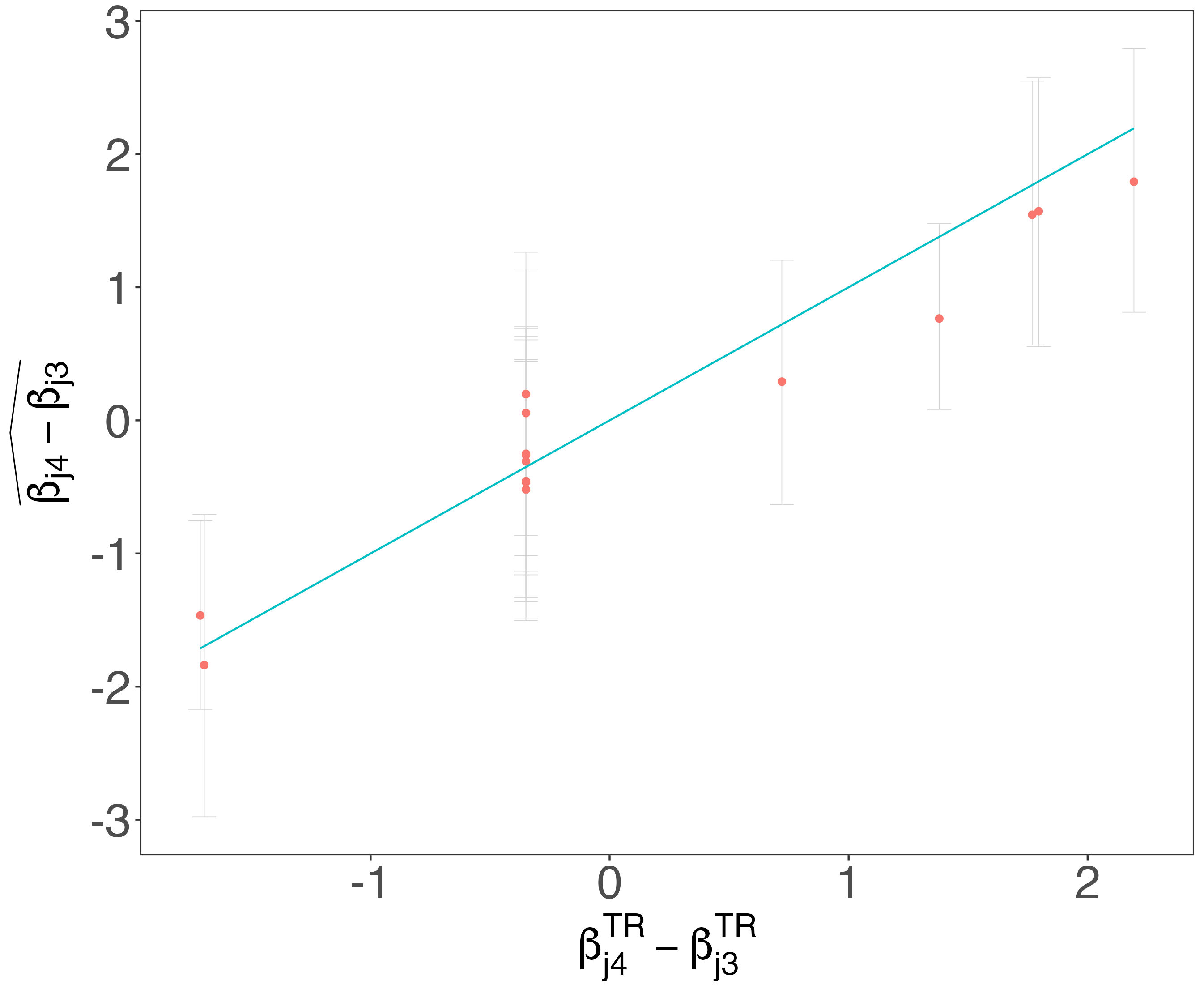} \\
   (a)  $\hat{\mu}_{ij}$ vs $\mu^\true_{ij}$ & (b) $\widehat{\beta_{j1}-\beta_{j2}}$ vs $\beta^\true_{j1}-\beta^\true_{j2}$ &(c) $\widehat{\beta_{j4}-\beta_{j3}}$ vs $\beta^\true_{j4}-\beta^\true_{j3}$ \\
   \vspace{0.02in}\\
   \includegraphics[width=.33\textwidth]{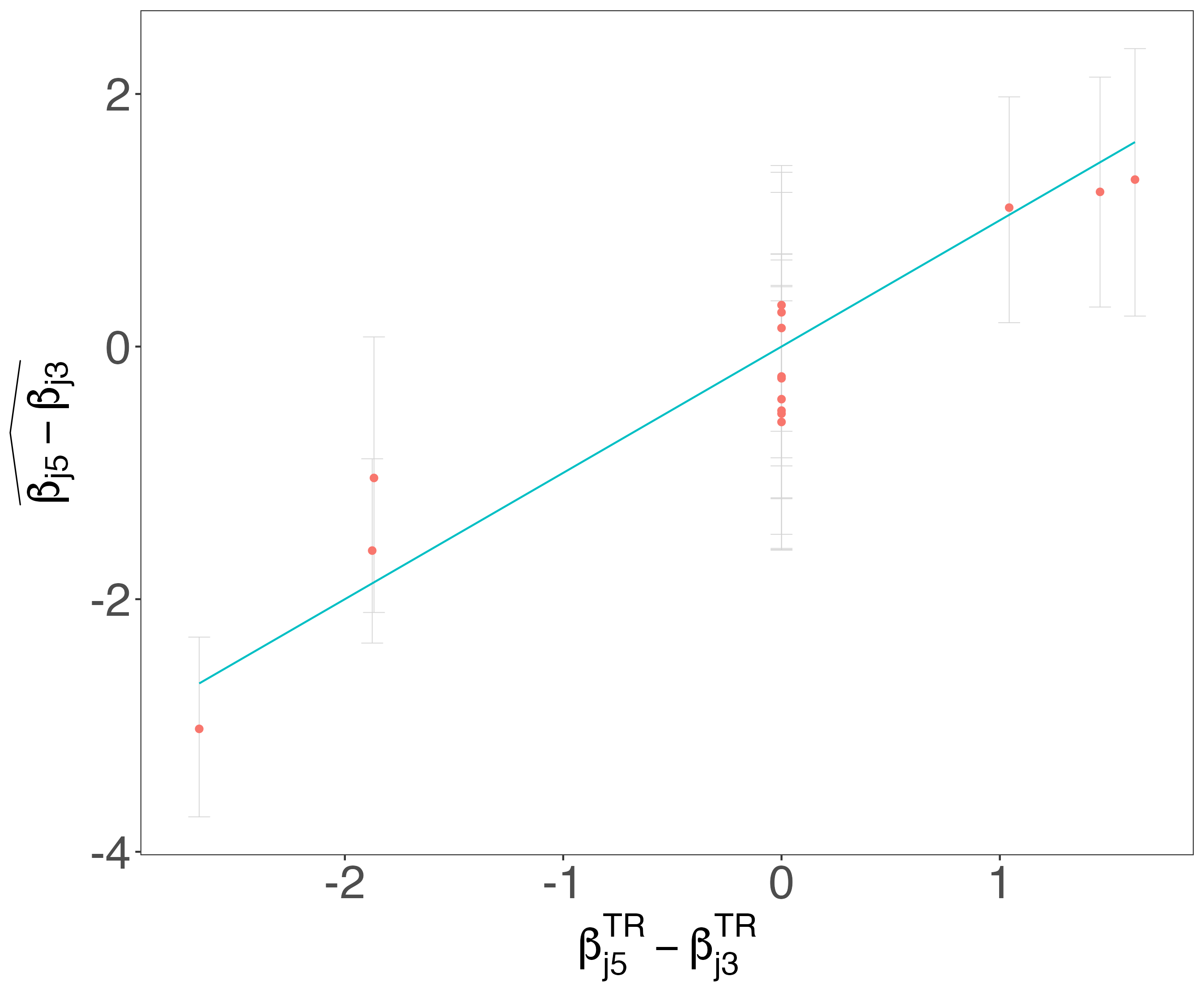}  & \includegraphics[width=.33\textwidth]{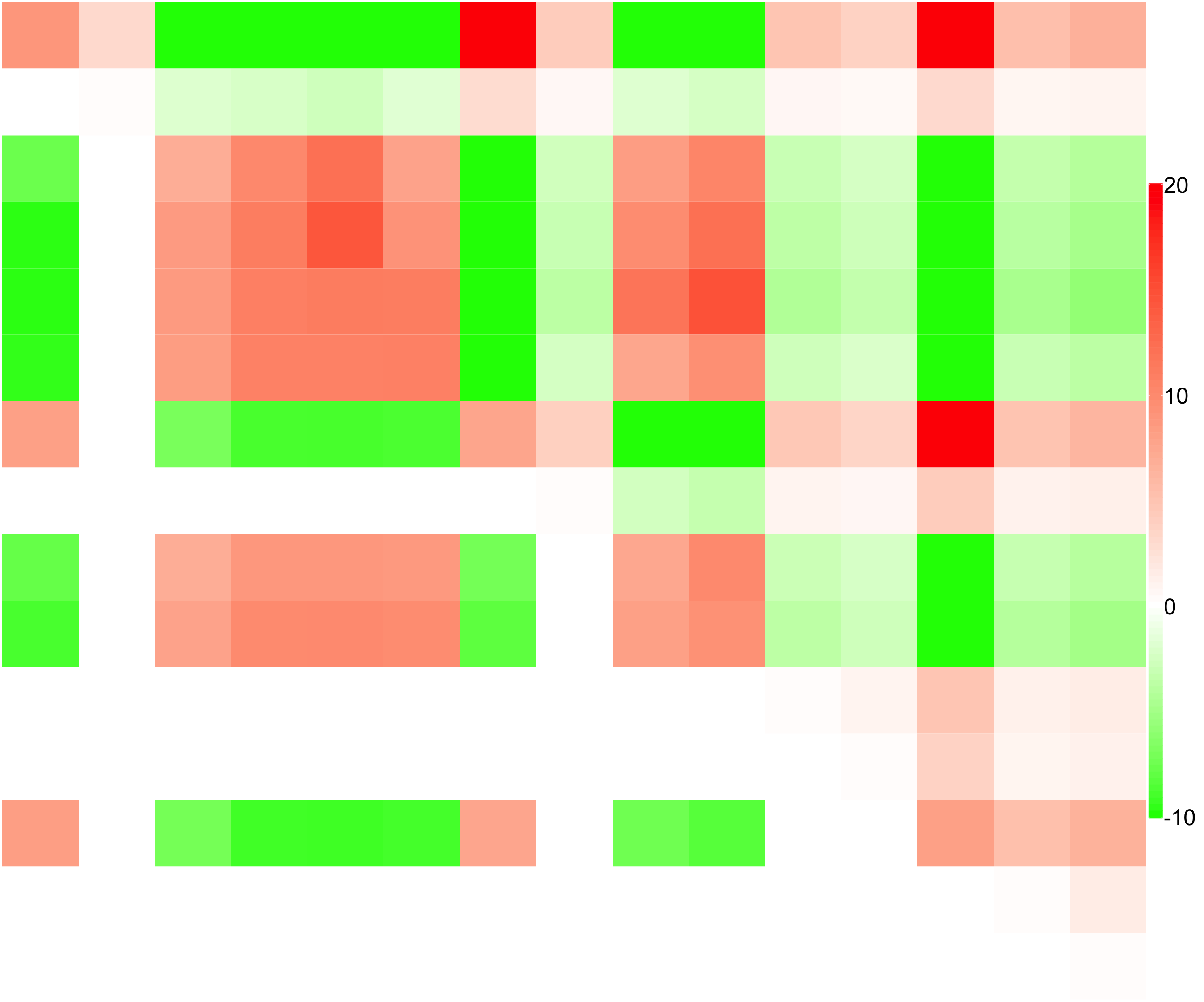}  & \includegraphics[width=.33\textwidth]{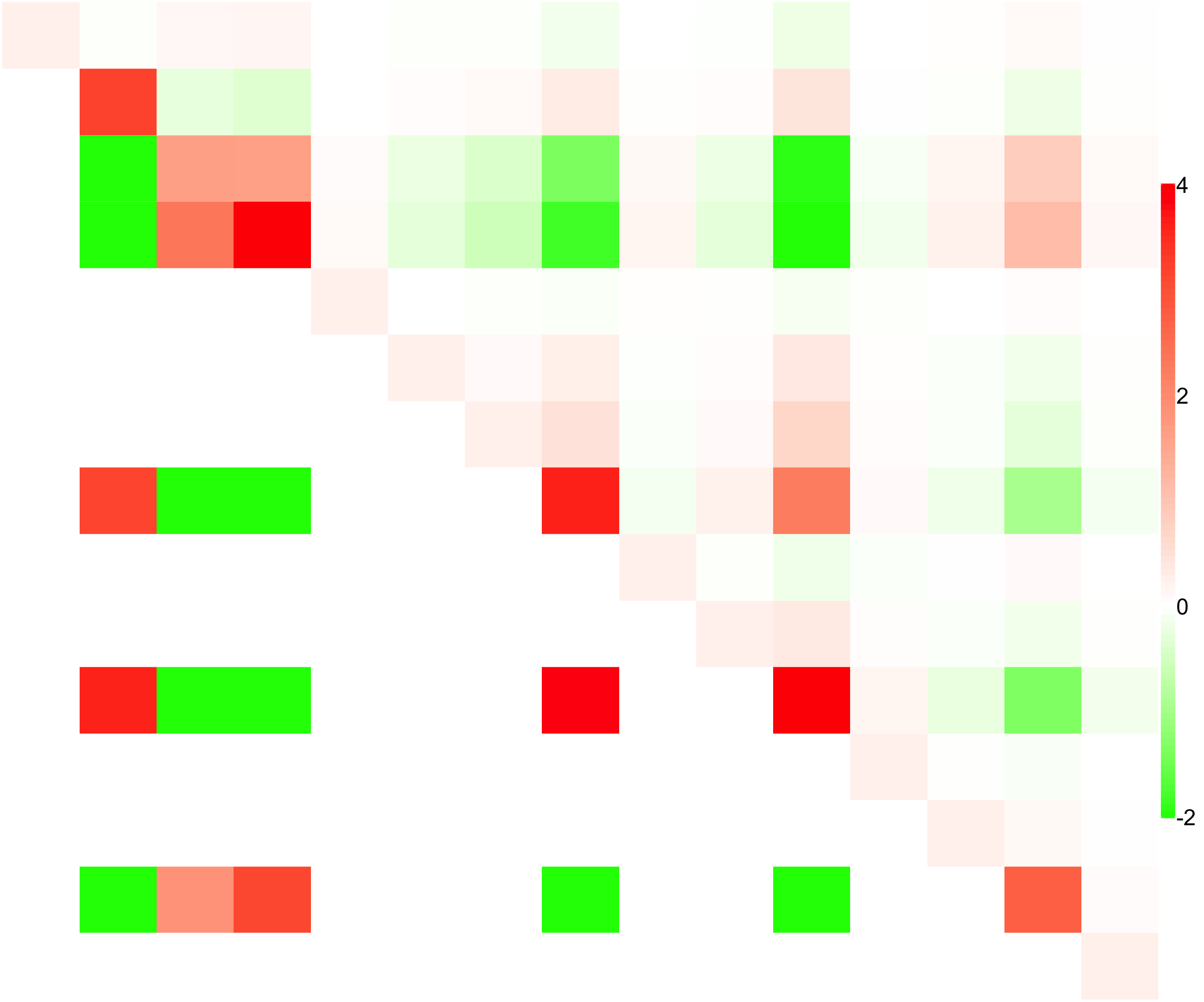} 
   \\
    (d) $\widehat{\beta_{j5}-\beta_{j3}}$ vs $\beta^\true_{j5}-\beta^\true_{j3}$  & (e) MOFA+: $\hat{\Sigma}(\bx)$ & (f) MOFA+: $\hat{\Sigma}(\bx^\prime)$ \\
  \end{tabular}
 \end{center}
 \vspace{-0.15in}
 \caption{[Simulation 1] The posterior median estimate of mean abundance $\mu_{ij}$ is plotted against the truth in panels (a). Panels (b)-(d) plot posterior point estimates of $\beta_{j1}-\beta_{j2}$, $\beta_{j4} -\beta_{j3}$ and $\beta_{j5} -\beta_{j3}$ against their true values. 
 The red dots represent posterior median estimates, and the grey vertical lines 95\% posterior credible interval estimates. Panel (e) and (f) illustrate the covariance estimate of MOFA+ for two arbitrarily selected conditions, $\bx=(1, 0, 0, 1)^\prime$ and $\bx^\prime=(1, 1, 0, 0)^\prime$.}
\label{fig:ch4-sim2-1-F2}
\end{figure}

Fig.~\ref{fig:ch4-sim2-1-F1}(a) presents a histogram of the differences $\hat{\Sigma}_{jj^\prime}(\bx_i) - \Sigma^\true_{jj^\prime}(\bx_i)$, $j \leq j^\prime$, across all samples. The differences are tightly centered around 0, indicating that the model provides accurate estimates of the covariance that varies with $\bx$. Figs.~\ref{fig:ch4-sim2-1-F1}(b) and (c) compare posterior median estimates of $\Sigma_{jj^\prime}(\bx)$ in the upper triangles to the true values $\Sigma^\true_{jj^\prime}(\bx)$ in the lower triangles for two arbitrarily selected conditions, $\bx=(1,0,0,1)^\prime$ and $\bx=(1,1,0,0)^\prime$, respectively. The true interaction patterns differ substantially across these conditions, both in terms of which features interact and the strength of their interactions. Comparing $\hat{\Sigma}_{jj^\prime}(\bx)$ to $\Sigma^\true_{jj^\prime}(\bx)$ shows that our method produces reasonable estimates for both conditions. $\hat{\Sigma}_{jj'}(\mathbf{x})$ for the remaining four conditions is illustrated in Supplementary Fig.~4(a)-(d).  Additional results for $\lambda(\mathbf{x}_i)$, $\tau_k$, and $\sigma^2$ are provided in Supplementary Fig.~3.

Fig.~\ref{fig:ch4-sim2-1-F2}(a) compares the posterior median estimates of $\mu_{ij}$ to their true values, with red dots representing the posterior medians and grey vertical lines representing 95\% credible intervals. Overall, the means of $Y^\star_{ij}$ are well estimated. Figs.~\ref{fig:ch4-sim2-1-F2}(b)-(d) illustrate the posterior median estimates of covariate effects on the mean abundance. Recall that $\beta_{jp}$ quantifies the change in mean abundance from its baseline, $\alpha_j + r_i$. The difference $\beta^\true_{j1} - \beta^\true_{j2}$ quantifies the effect of the binary covariate, while the differences $\beta^\true_{j4} - \beta^\true_{j3}$, $\beta^\true_{j5} - \beta^\true_{j3}$, and $\beta^\true_{j5} - \beta^\true_{j4}$ quantify the pairwise effects of the ternary covariate on $\mu_{ij}$. Panels (b)-(d) show that the covariate effects on the mean abundance are well estimated. A comparison of $\widehat{\beta_{j5} - \beta_{j4}}$ to its true value is provided in Supplementary Fig.~3.

We compare our method to MOFA+ \citep{argelaguet2020mofa}, a multi-omics factor analysis framework that provides joint association estimates across heterogeneous datasets. MOFA+ extends the original MOFA model \citep{Argelaguet2018} by modeling both shared and modality-specific sources of variability across multiple data views. We treat each category as a different view and apply MOFA+ to the full dataset. The resulting covariance estimates are presented in Fig.~\ref{fig:ch4-sim2-1-F2}(e)-(f) and Supplementary Fig.~4 (e)-(h). MOFA+ performs well with categorical covariates but misses some dependencies at certain levels.  
We also applied SparCC and CCLasso separately to each dataset corresponding to the individual categorical levels, since these methods do not incorporate $\bx$. Supplementary Fig.~4 compares their correlation estimates to the true values. We report the RMSE of the estimated correlations $\hat{\rho}_{jk}(\bx) = \widehat{\Sigma(\bx)_{jk}/\sqrt{\Sigma(\bx)_{jj}\Sigma(\bx)_{kk}}}$ ($j<k$) for all methods in Table~\ref{tab:rmse-comparison}. Compared to the other methods, our model more accurately captures the structured, covariate-varying covariance patterns, demonstrating its superior ability to recover feature interactions that vary with covariates.

\begin{table}[t!]
\centering
\caption{RMSE of the estimated correlations $\hat{\rho}_{jk}(j<k)$ under all methods across three simulation settings.}
\label{tab:rmse-comparison}
\begin{tabular}{lcccc}
\toprule
\textbf{Simulation} & \textbf{Our Model} & \textbf{MOFA+} & \textbf{SparCC} & \textbf{CCLasso} \\
\midrule
1 & 0.069 & 0.602 & 0.832 & 0.465 \\
2 & 0.058 & 0.336 & 0.265 & 0.242 \\
3 & 0.126 & 0.357 & 0.267 & 0.231 \\
\bottomrule
\end{tabular}
\end{table}

\subsection{Simulation 2}\label{sec:ch4-sim2}
Simulation 2 considers a dataset with repeated samples that typically has large inter-subject variability.  Suppose we have multiple samples from each subject in a set. We let $s_i \in \{1, \ldots, S\}$ denote the subject from which sample $i$ is taken. To accommodate inter-subject variability, we extend the model in a fashion similar to that in \cite{zhang2023}; we let the normalized mean abundance indexed by subjects, that is, $\alpha_{s_i, j}$ and assumed the following model for $\balpha_{s_i}=(\alpha_{s_i1},\ldots, \alpha_{s_iJ})^\prime$;
\begin{eqnarray}
\begin{aligned}
\balpha_{s_i} \mid G  \iid& G(\balpha), \hspace{0.01in} s_i \in \{1, \ldots, S\},\\
G(\balpha) = \prod_{j=1}^{J}G_{j}(\alpha_{j}) = \prod_{j=1}^{J} &
\left[\sum_{l=1}^{\infty} \psi_{l}^\alpha
\left\{ \omega_{l}^\alpha\delta_{\xi_{jl}^\alpha}+(1-\omega_{l}^\alpha)\delta_{\left(\frac{\nu_{j}^\alpha-\omega_{l}^\alpha\xi_{jl}^\alpha}{1-\omega_{l}^\alpha}\right)}\right\}\right].
\label{eq:ch4-prior-balpsij} 
\end{aligned}
\end{eqnarray}
That is, the baseline abundance of an OTU is shared by all samples from a subject, and the prior distribution of $\balpha_{s}$ has different mixing locations $\xi^\alpha_{jl}$ for each OTU to accommodate variability in the baseline abundance among OTUs.

\begin{figure}[!t]
  \begin{center}
\begin{tabular}{ccc}
   \includegraphics[width=.32\textwidth]{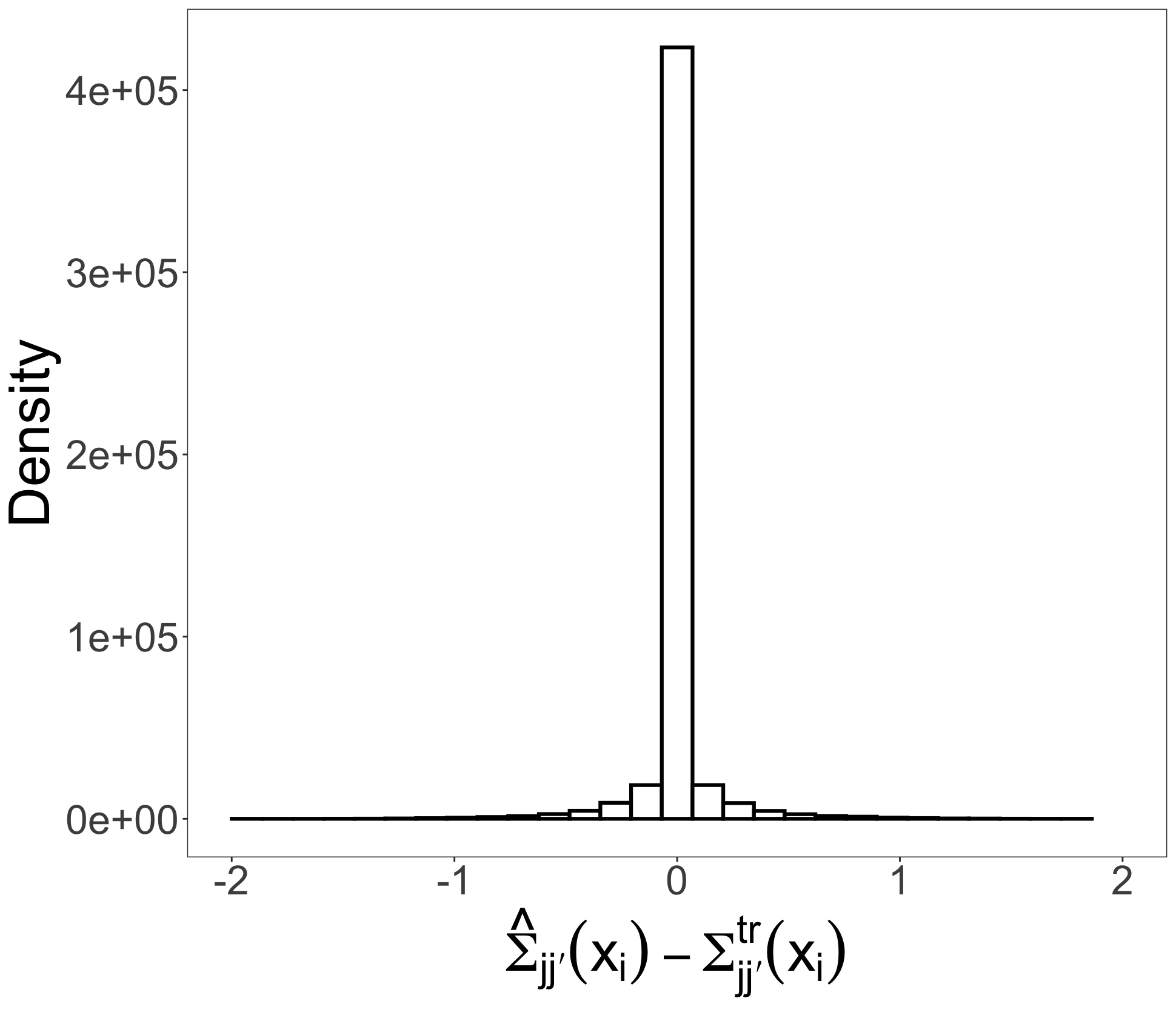} & 
   \includegraphics[width=.33\textwidth]{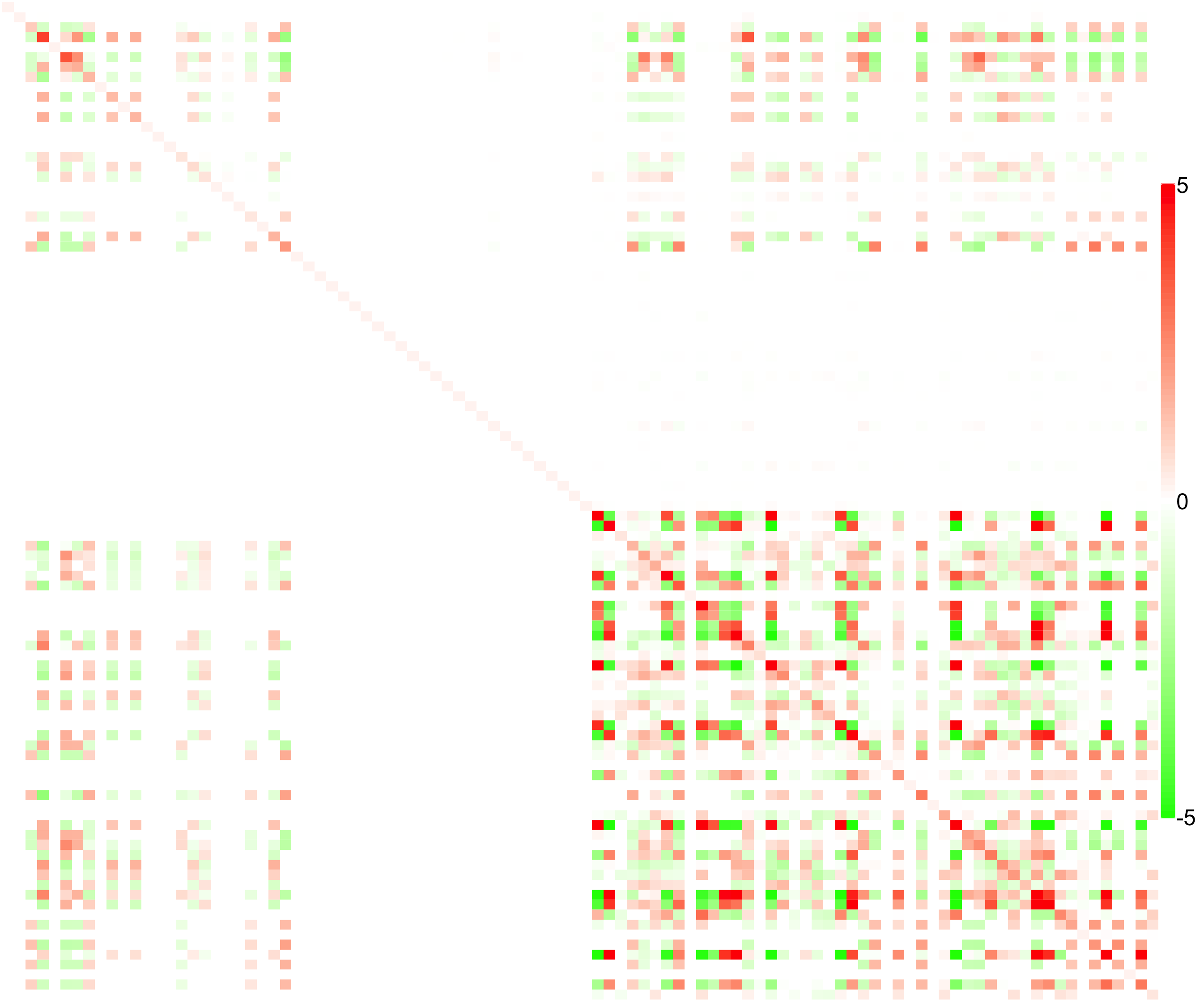} &
   \includegraphics[width=.33\textwidth]{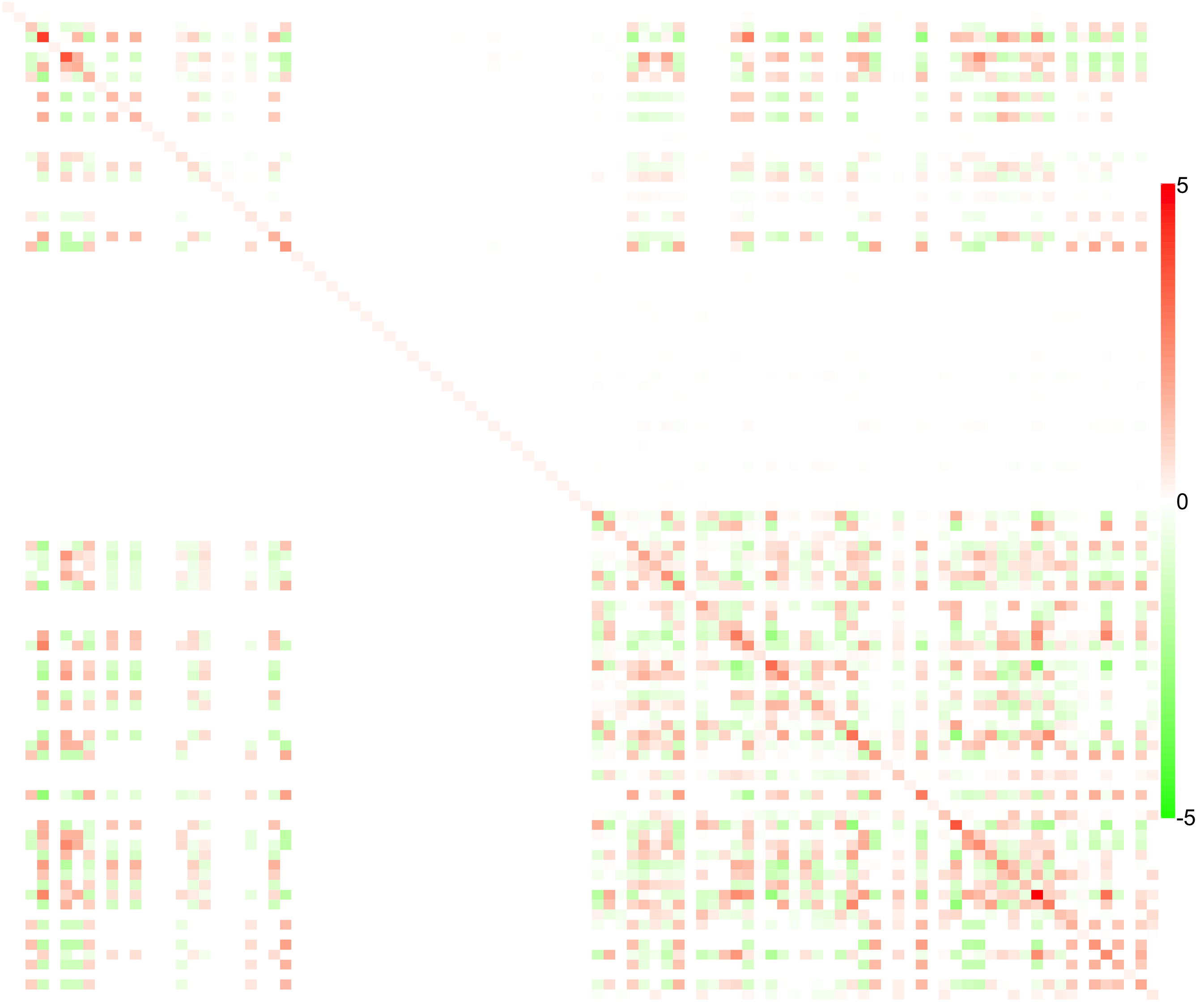} \\
   (a) $\hat\Sigma_{jj^\prime}(\bx)-\Sigma^{\true}_{jj^\prime}(\bx)$ & (b) $\hat\Sigma(\bx_2)$ vs $\Sigma^{\true}(\bx_2)$ & (c) $\hat\Sigma(\bx_{27})$ vs $\Sigma^{\true}(\bx_{27})$
  \end{tabular}
 \end{center}
 \vspace{-0.15in}
 \caption{[Simulation 2] Panel (a) has a histogram of differences between $\hat\Sigma_{jj^\prime}(\bx_i)$ and $\Sigma^{\true}_{jj^\prime}(\bx_i)$ of all samples. In (b), the lower left and upper right triangles of the heatmap illustrate true values $\Sigma^{\true}_{jj^\prime}$ and their posterior estimates of correlations $\hat{\Sigma}_{jj^\prime}$, respectively. Two samples, samples 2 and 27, from subject 2, are arbitrarily chosen for illustration. Their covariates are $\bx_{2}=(1,-1.23), \bx_{27}=(0,-1.23)$. } 
\label{fig:ch4-sim1-F1}
\end{figure}

To generate a dataset, we let $S=25$ subjects and $J=100$ OTUs. We included a continuous subject covariate $x^c_{s_i}\iid \Nor(0,1)$ and a binary sample covariate, $x^d_i\in\{0,1\}$, that represents two experimental conditions. Assuming that a sample is obtained from each condition for a subject, we had $N=50$ samples.  Adapting the common factor model in \cite{de2019multi, de2021bayesian}, we let
$
\Sigma^\true(\bx)=\Lambda^\true_0\Lambda^{\true,\prime}_0+\Lambda^\true(\bx)\Lambda^{\true,\prime}(\bx)+\sigma^{2, \true} \uI_J, 
$
where $\Lambda^\true_0$ is $J \times K_0$ matrix of common factor loadings with $K_0=2$, and $\Lambda^\true_0\Lambda^{\true,\prime}_0$ is a baseline covariance. We simulated random variables from $\Nor(0, 1)$ and shifted them away from zero by 1/2 for $\lambda^\true_{0,jk}$ of OTUs 1-25 and 51-100 to ensure that those OTUs have baseline interactions. We let $\lambda^\true_{0,jk}$ = 0 for the remaining OTUs. 
We next let  $\Lambda^\true(\bx)$ a $J \times K_1$ factor loading matrix with $K_1=3$ and let $\lambda^\true_{1, jk}(\bx)=q^\true_{jk}\bm f^{\true, \prime}_k\bx$. We then simulated random variables from $\Nor(0, 1)$, shifted them away from zero by 1/2, and set them to be $f^\true_{jk}$ for OTUs 51-100 and $f^\true_{jk}=0$ for the remaining OTUs. And we have $f^\true_{kp}\iid \unif(-1,1)$ and $\sigma^{2,\true}=0.5^2$. 
Under this design, interactions among OTUs 1-25 remain constant across covariates, while among OTUs 51-100 change with the covariates. Furthermore, OTUs 26-50 do not interact with the other OTUs.  The covariance matrix corresponding to $\Sigma^\true(\bx_i)$ is illustrated in the lower triangle of Fig~\ref{fig:ch4-sim1-F1}(b) and (c) for two selected samples.
Excess zeros and large inter-subject variability are commonly observed in microbiome data, and we reflect those in the simulated dataset.  For the normalized abundance level, we first set $\xi^{\alpha, \true}_{j1} = -5$, $\xi^{\alpha, \true}_{j2} \sim \Nor(2.5,0.5)$ and $\xi^{\alpha, \true}_{j3} \sim \Nor(5, 0.5)$ and simulated $\bm\psi^\true_{j} = (\psi^\true_{j1}, \psi^\true_{j2}, \psi^\true_{j3})\sim \Dir(30, 40, 30)$. The three values, $\xi^{\alpha, \true}_{jl}$, $l=1$, 2 and 3, represent zero, small and large counts, respectively.  We then let $\alpha^\true_{sj}=\xi^{\alpha, \true}_{jl}$ with probability $\psi^\true_{jl}$ for $s \in \{1, \ldots, S\}$. 
We next simulated size factors $r^{\true}_{i}\iid \unif(0,2)$ and regression coefficients $\beta^{\true}_{jp}$ similar to Simulation 1. 
We let $\bmu^{\true}(\bx_i) = r^{\true}_i\bm{\mbox{1}}_J + \bm\alpha^{\true}_{s_i}+\bbet^\prime\bx_i$, and sample count vector  $\bY_{i}$ from $\lfloor \logN_J(\bmu^{\true}(\bx_i), \Sigma^{\true}(\bx_i)) \rfloor$.
Under this setup, approximately 30.76\% of $Y_{ij}$'s are 0. We specified the hyper-parameters values similar to Simulation 1 with $K=7$. We ran MCMC for $160,000$ iterations and discarded the first half for burn-in. It took 23 hours on an Apple M1 chip laptop.

\begin{figure}[!t]
  \begin{center}
\begin{tabular}{ccc}
   \includegraphics[width=.31\textwidth]{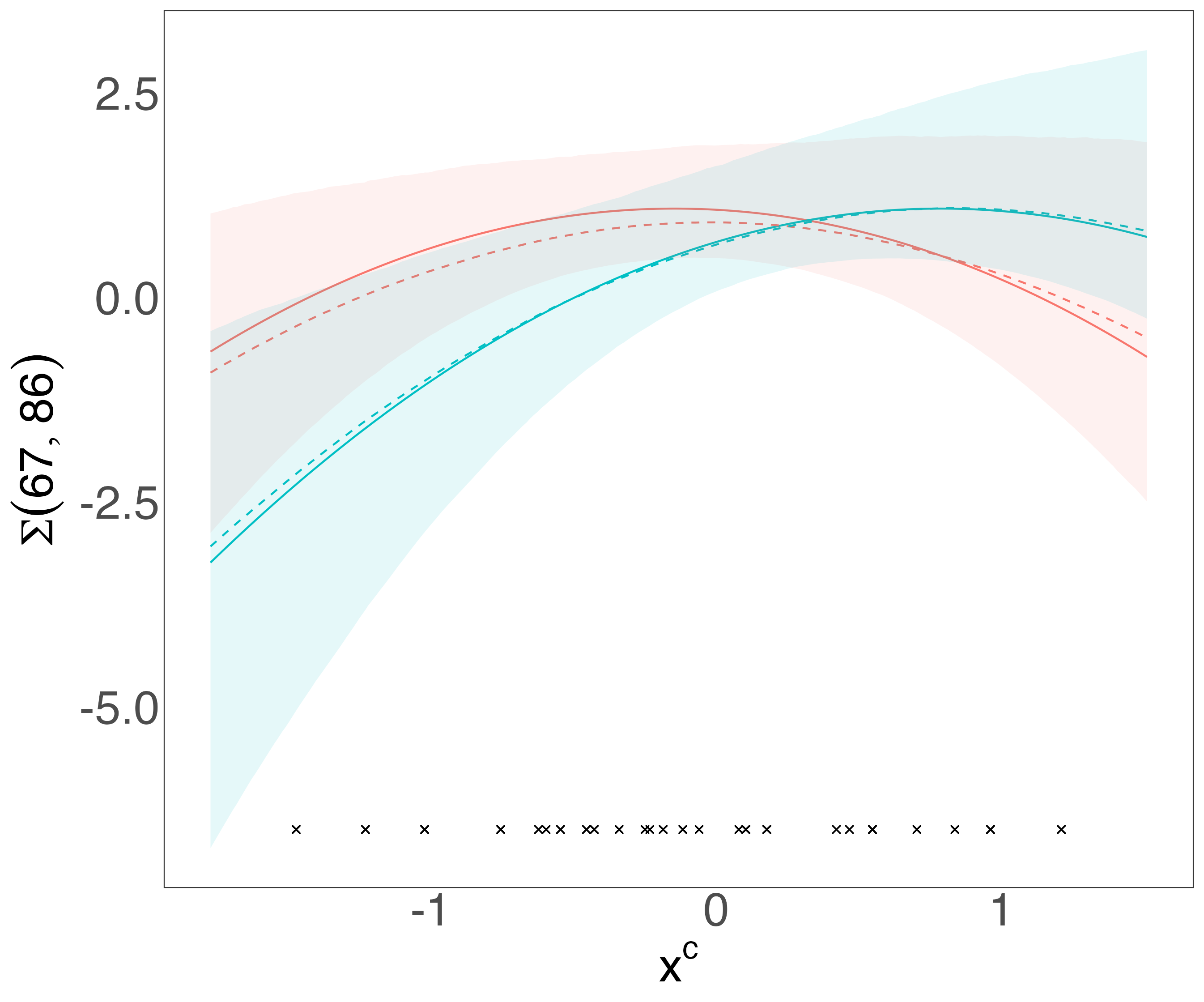} & 
   \includegraphics[width=.31\textwidth]{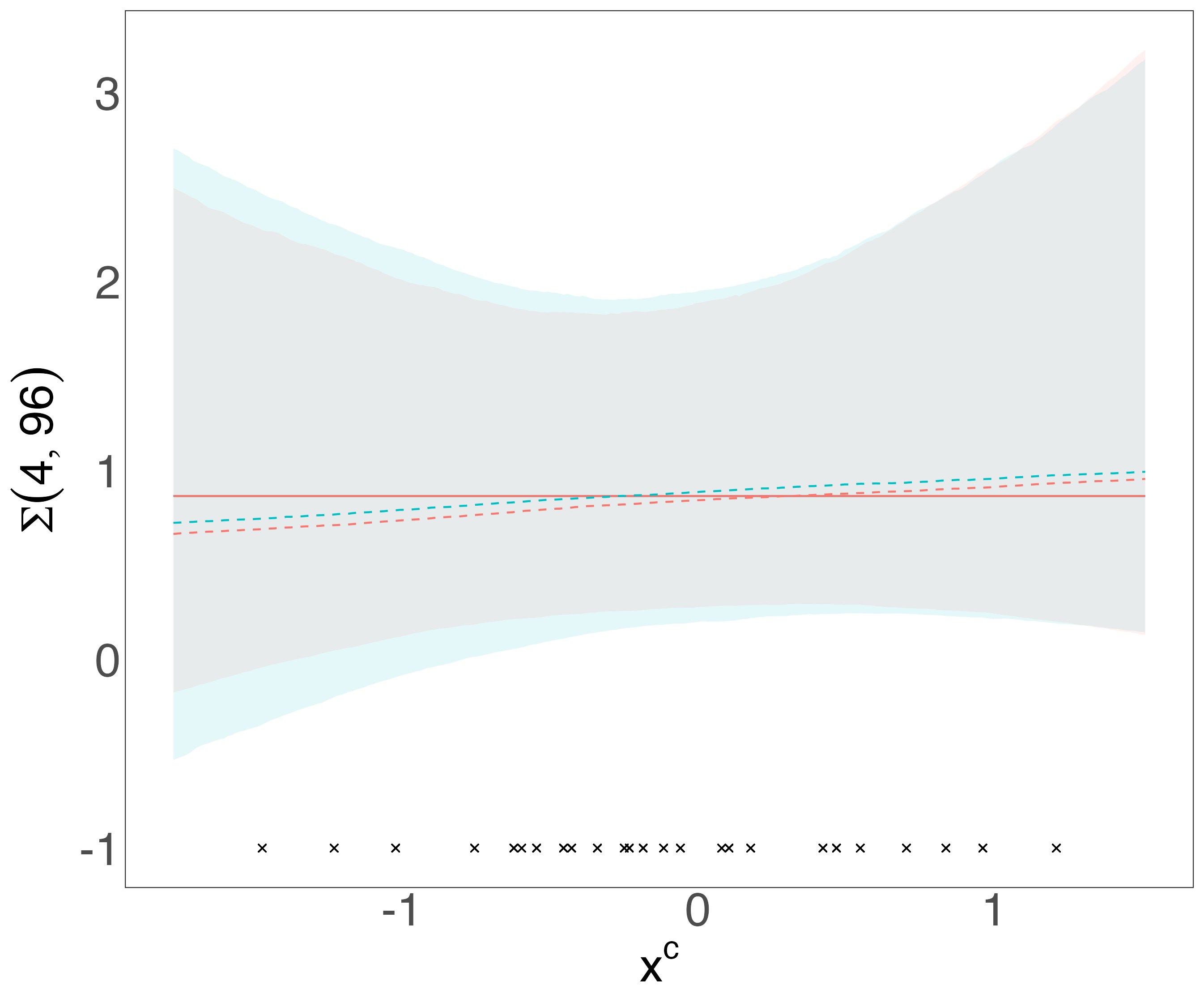} &
   \includegraphics[width=.31\textwidth]{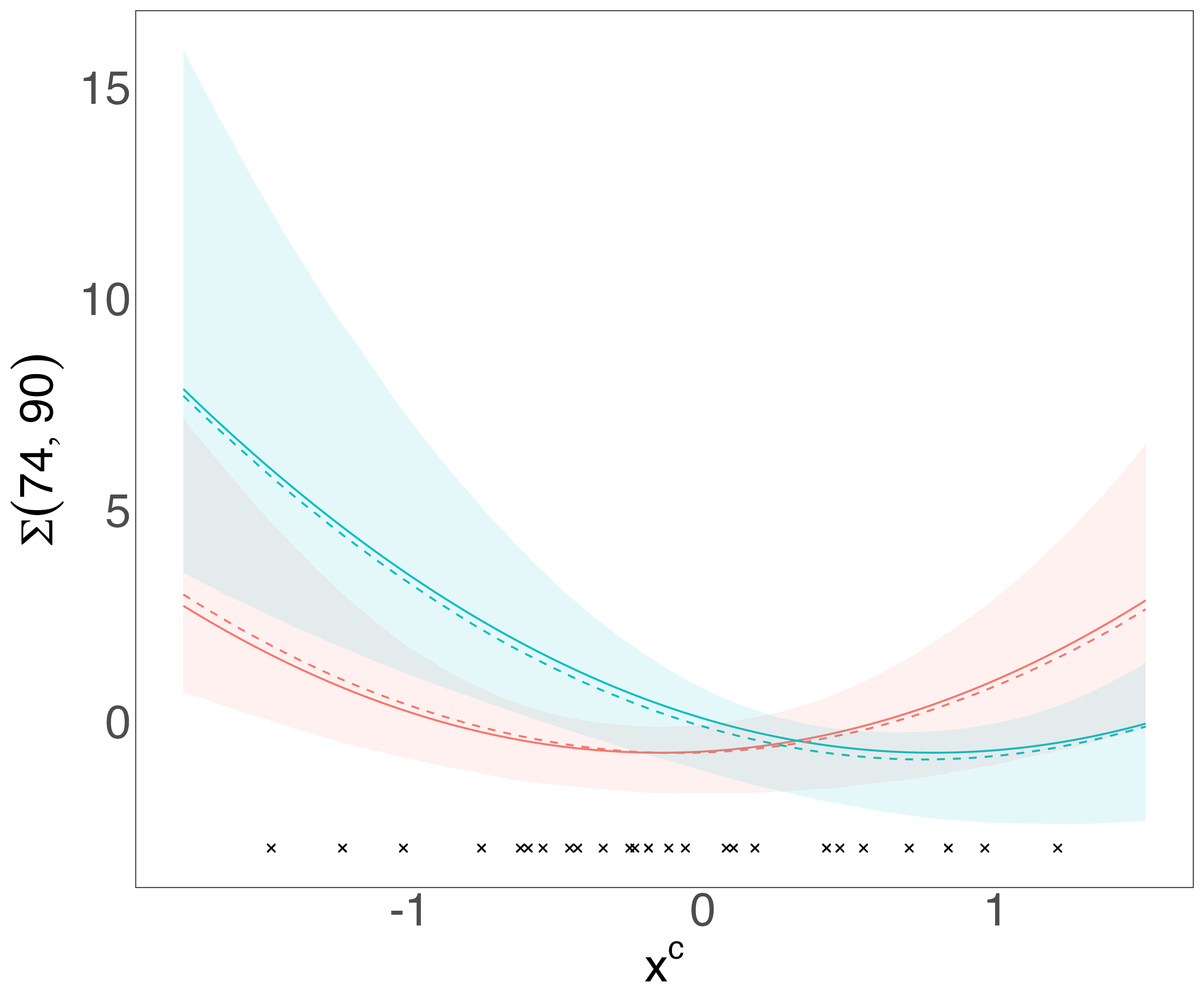} \\
   (a)  OTUs 67 \& 86 & 
   (b)  OTUs 4 \& 96  & 
   (c)  OTUs 74 \& 90 
  \end{tabular}
 \end{center}
 \vspace{-0.15in}
 \caption{[Simulation 2] Scatter plots of $\Sigma_{jj^\prime}(\bx)$ (dashed line) and $\Sigma^\true_{jj^\prime}(\bx)$ (solid line) are plotted for three arbitrarily chosen OTU pairs, OTUs 67 and 86 in panel (a), OTUs 4 and 96 in panel (b), and OTUs 74 and 90 in panel (c). Crosses are observed values of the continuous covariate $x_c$.  The red and blue colors are for $x^d=0$ and 1, respectively.  The shades represent pointwise 95\% posterior credible interval estimates.}
\label{fig:ch4-sim1-F2}
\end{figure}

\begin{figure}[!t]
  \begin{center}
\begin{tabular}{cc}
   \includegraphics[width=.475\textwidth]{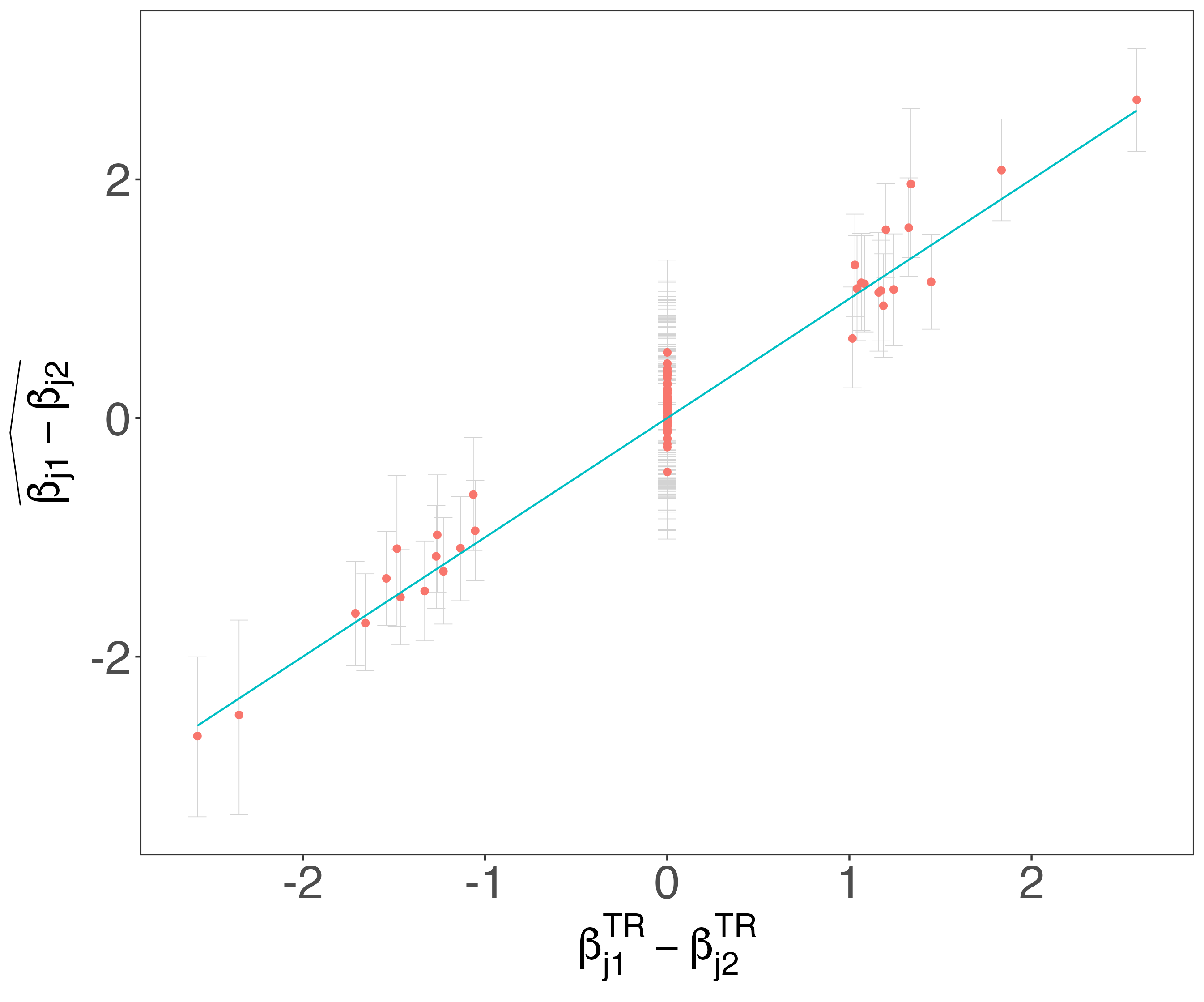} & 
   \includegraphics[width=.475\textwidth]{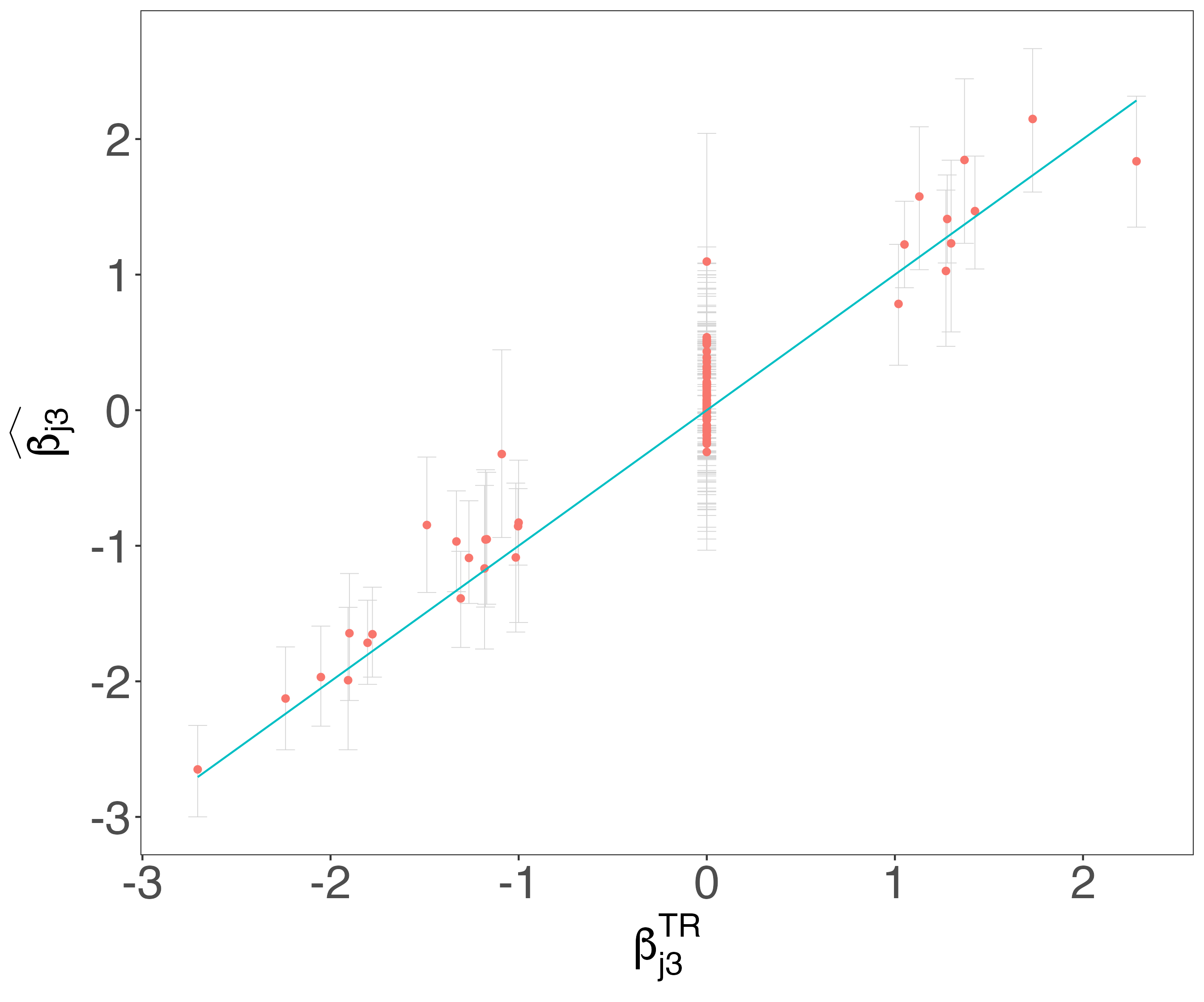} \\
   (a) $\widehat{\beta_{j1}-\beta_{j2}}$ versus $\beta_{j1}^{\true}-\beta_{j2}^{\true}$ &
   (b) $\hat{\beta}_{j3}$ versus $\beta_{j3}^{\true}$ 
  \end{tabular}
 \end{center}
 \vspace{-0.15in}
 \caption{[Simulation 2] The posterior estimates of the effects of the binary and continuous covariates on abundance are plotted in panels (a) and (b), respectively. The dots represent the posterior median estimates, while the vertical lines indicate their corresponding 95\% credible interval estimates.}
\label{fig:ch4-sim1-F3}
\end{figure}

Fig~\ref{fig:ch4-sim1-F1}(a) plots the differences $\hat{\Sigma}_{jj^\prime}(\bx_i) - \Sigma^\true_{jj^\prime}(\bx_i)$ of all samples, $j \leq j^\prime$. It shows that the extended model captures covariate varying feature interactions reasonably well even when excess zeros are present and counts greatly vary across subjects.  Fig~\ref{fig:ch4-sim1-F1}(b) and (c) compare the posterior median estimates of $\Sigma(\bx)$ to its truth. Their $\bx$'s have the same value of the continuous covariate, $x^c=-1.23$, but the discrete covariate $x_d$ takes the value of 0 and 1, respectively.  
Fig~\ref{fig:ch4-sim1-F2} presents the posterior estimates of covariance for some selected pairs of OTUs by varying the values of $x^c$ and $x^d$.  The solid and dashed lines are the truth and posterior estimates, and the red and blue colors correspond to $x^d=0$ and 1, respectively.  The shaded area represents point-wise 95\% posterior credible interval estimates. The crosses at the bottom of the plots are the observed values of $x^c$.   
Figs~\ref{fig:ch4-sim1-F1} and~\ref{fig:ch4-sim1-F2} show that our model successfully identifies inactive OTUs, accurately estimates the baseline covariance, and captures the covariate-varying interrelationship structure among OTUs even when the sample size is smaller than the number of OTUs. Fig~\ref{fig:ch4-sim1-F3} illustrates the accuracy of our model in estimating covariate effects on microbial abundances under Simulation~2. The estimates align closely with the diagonal line, indicating that the proposed model accurately recovers both binary and continuous covariate effects. 

More details on the posterior median estimates of the loading matrices, shrinkage parameters, and mean abundances are provided in Supp Fig 5. The results indicate that the model effectively selects relevant factors and achieves high overall estimation accuracy. We also compared our method to comparators in Supp Fig 6 and compute RMSE shown in Table~\ref{tab:rmse-comparison}. Due to the continuous covariate, all three comparators, including MOFA+, cannot accommodate covariates and therefore produce a single global estimate shared across samples.

\subsection{Simulation 3}\label{sec:ch4-sim3}
\begin{figure}[!t]
  \begin{center}
\begin{tabular}{ccc}
   \includegraphics[width=.31\textwidth]{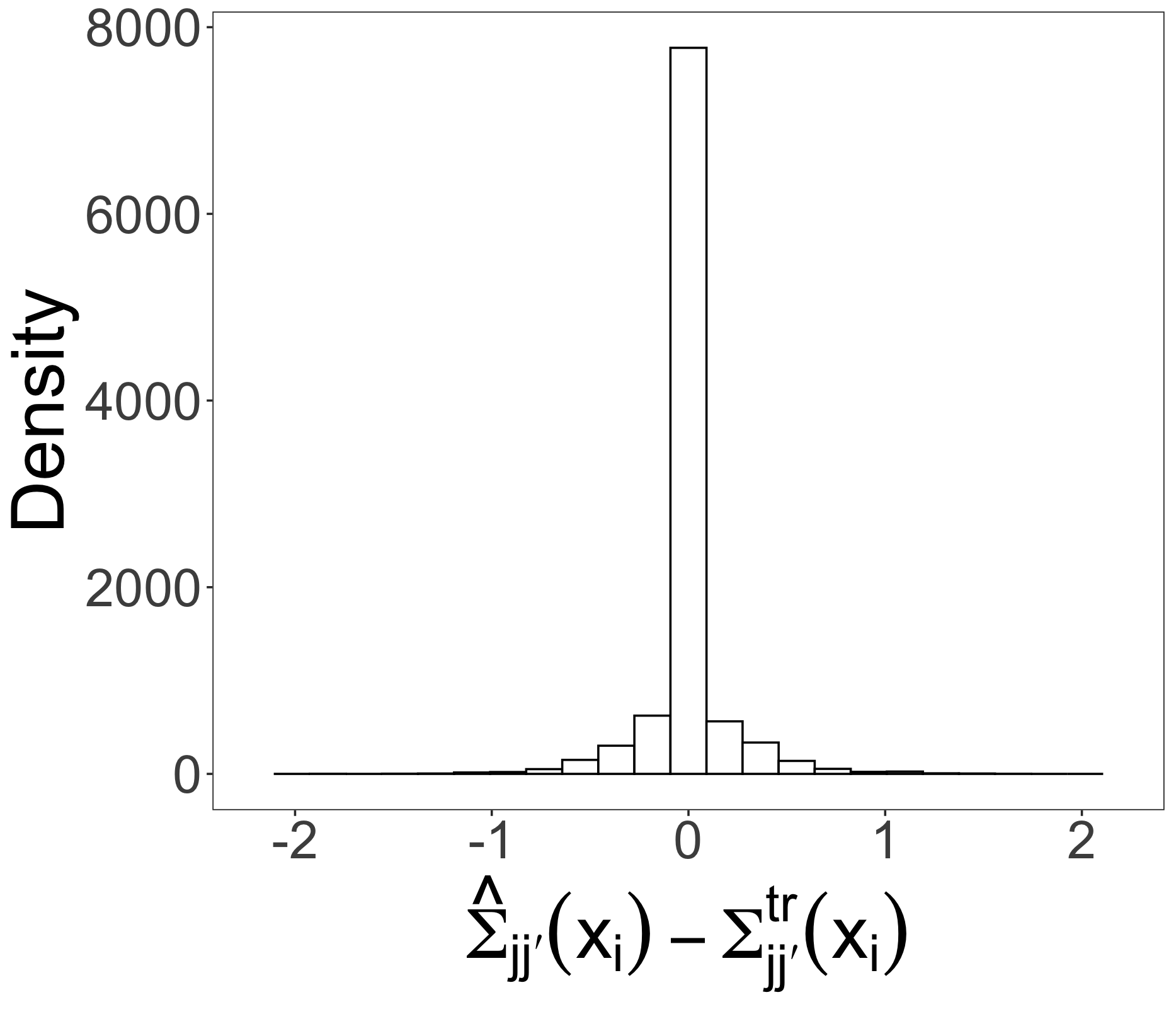} & 
   \includegraphics[width=.31\textwidth]{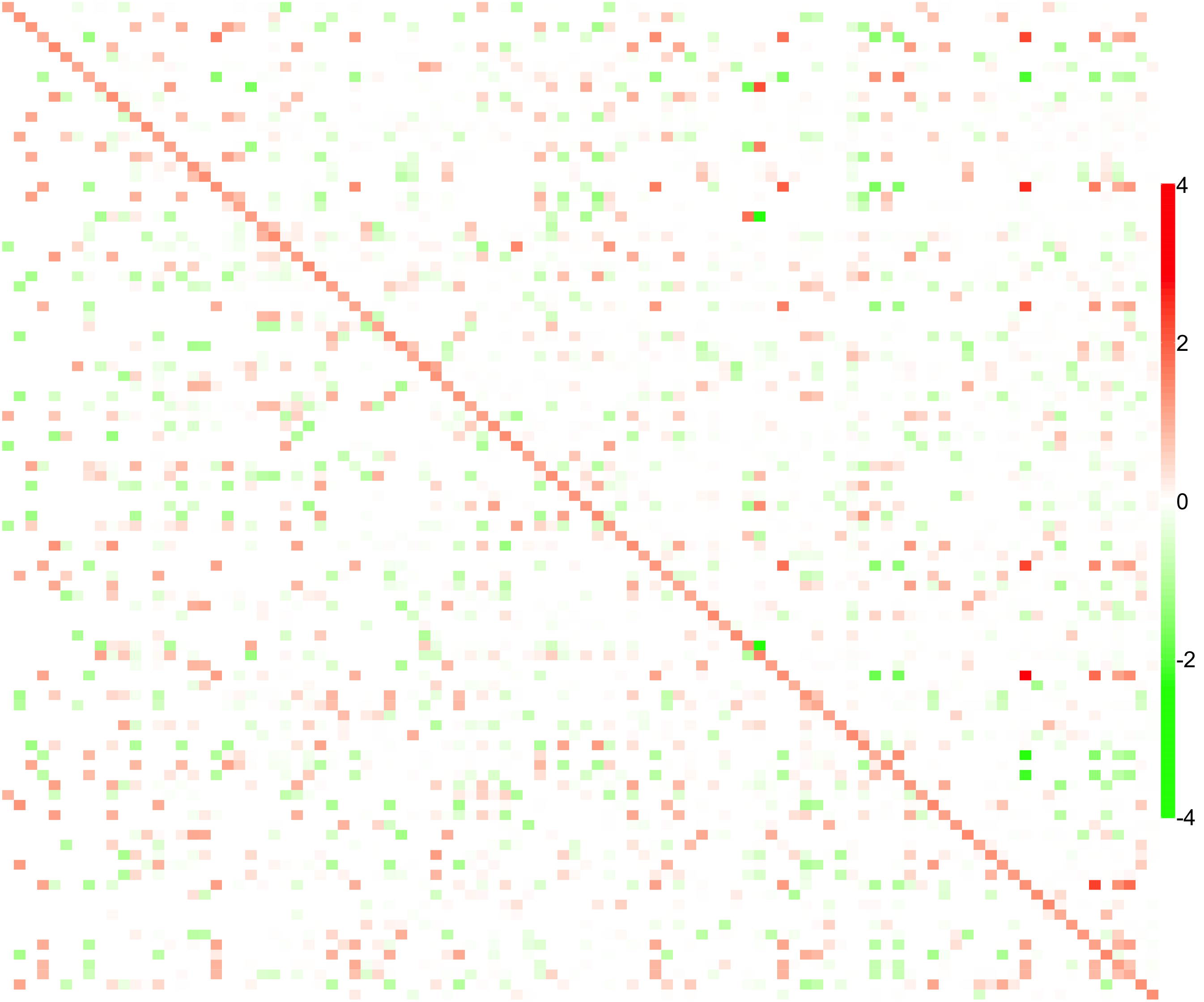} &
   \includegraphics[width=.31\textwidth]{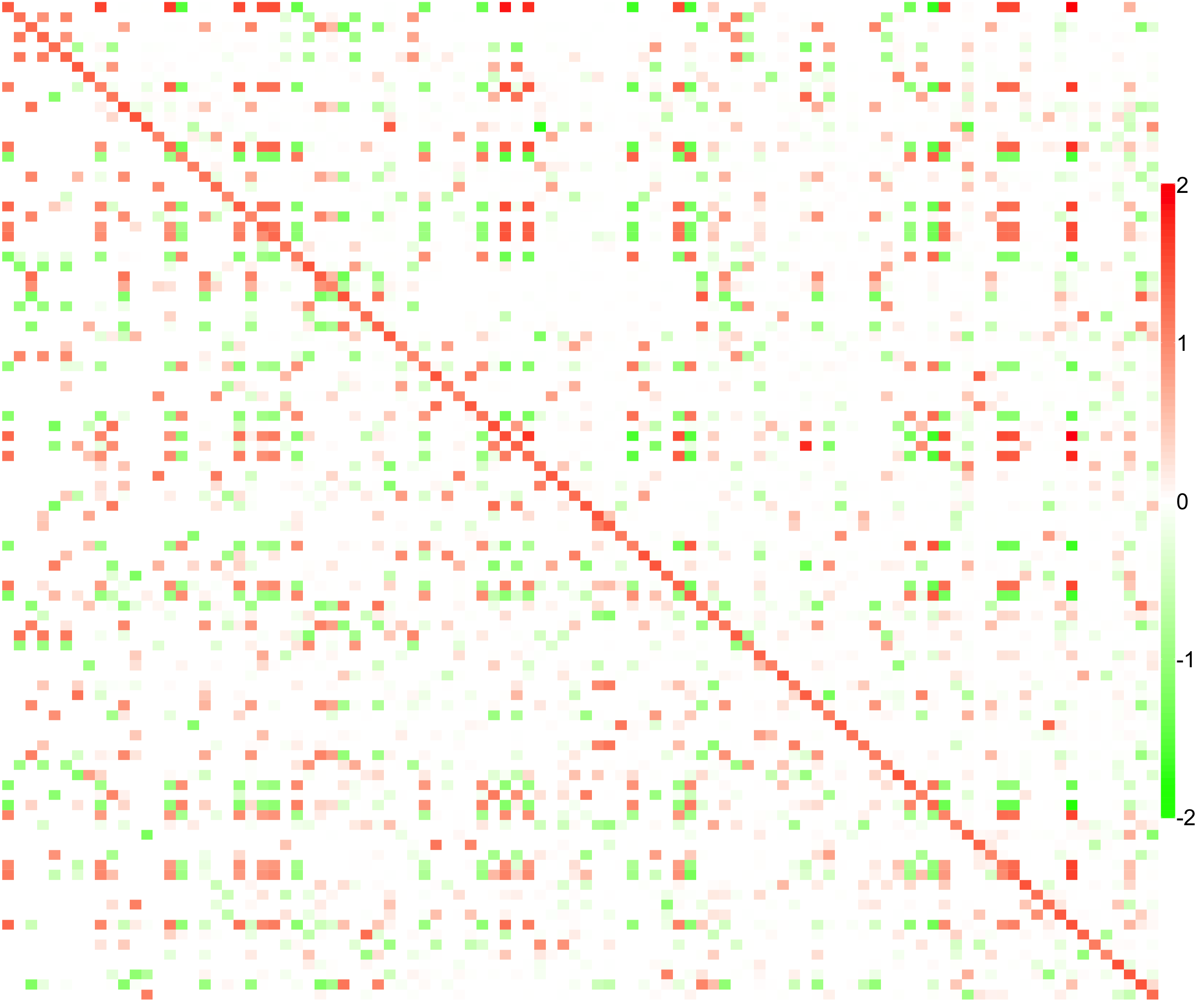} \\
   (a) $\hat\Sigma_{jj^\prime}-\Sigma^{\true}_{jj^\prime}$ & (b) $\hat\Sigma_{0}$ vs $\Sigma^{\true}_{0}$ & (c) $\hat\Sigma_{1}$ vs $\Sigma^{\true}_{1}$
  \end{tabular}
 \end{center}
 \vspace{-0.15in}
 \caption{[Simulation 3] Panel (a) presents a histogram of the differences between $\hat\Sigma_{jj^\prime}(x_d)$ and $\Sigma^{\true}_{jj^\prime}(x_d)$, for $j \leq j^\prime$ and $x_d \in \{0, 1\}$, where $\hat\Sigma_{jj^\prime}(x_d)$ are the posterior median estimates of $\Sigma_{jj^\prime}(x_d)$. Panels (b) and (c) compare $\hat{\Sigma}(x_d)$ to their true values for each $x_d$.} 
\label{fig:ch4-sim2-F1}
\end{figure}

For Simulation 3, we let the covariance matrix arbitrarily vary with a binary covariate, $x^d \in \{0, 1\}$ that represents two experimental conditions.  Specifically, arbitrary covariance matrices were generated using the vine method in \cite{LEWANDOWSKI20091989}, separately for each value of $x^d$, as follows; We simulated partial correlations from linearly transformed $\Be(1,1)$ distribution over the interval of $(-1,1)$. To encourage sparsity in $\Sigma^\true(\bx)$, we set the partial correlations below 0.8 to 0 and generated a correlation matrix, $\rho^\true(x^d)$ using their recursive formula. We then sampled $\sigma^{2,\true}_j(x^d)$ independently from $\unif(1, 1.5)$ and let $\Sigma_{jj^\prime}^\true(x^d)=\sigma^{2,\true}_j(x^d)\sigma^{2,\true}_{j^\prime}(x^d)\rho_{jj^\prime}^\true(x^d)$. $\Sigma^\true(0)$ and $\Sigma^\true(1)$ are shown in the lower triangle of Fig~\ref{fig:ch4-sim2-F1}(b) and (c), respectively. We kept the rest of the simulation setup the same as in Simulation 2, with $S=25$ subjects, $N=50$ samples, and $J=100$ OTUs. We also used the same fixed hyperparameter values as in Simulation 2 but choose $K= 25$ empirically. Following \S~\ref{sec:ch4-mcmc}, we perform PCA on clr transformed data and plotted the screen plot of eigenvalues in Supp Fig 5. Furthermore, we examined the prior sensitivity analysis by varying the value of $K$ and found that the covariate-varying covariance matrix estimates remain mostly unchanged for large enough values of $K$. More detailed results are summarized in Supp Fig 7. Examination of the MCMC simulation using traceplots indicated no evidence of convergence or mixing problems. It takes 16 hours on Apple M1Max. 

\begin{figure}[!t]
  \begin{center}
\begin{tabular}{cc}
   \includegraphics[width=.5\textwidth]{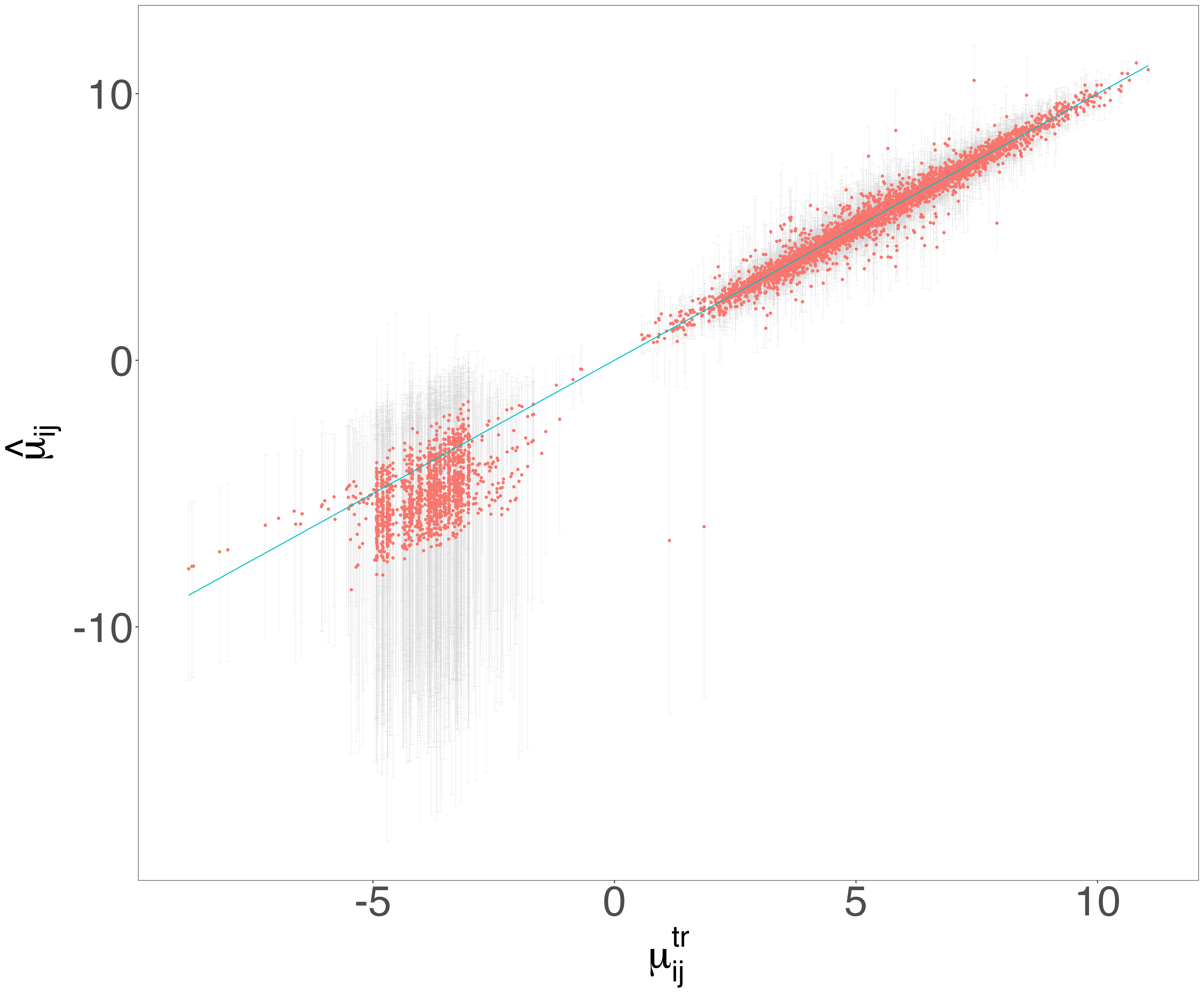} & 
   \includegraphics[width=.5\textwidth]{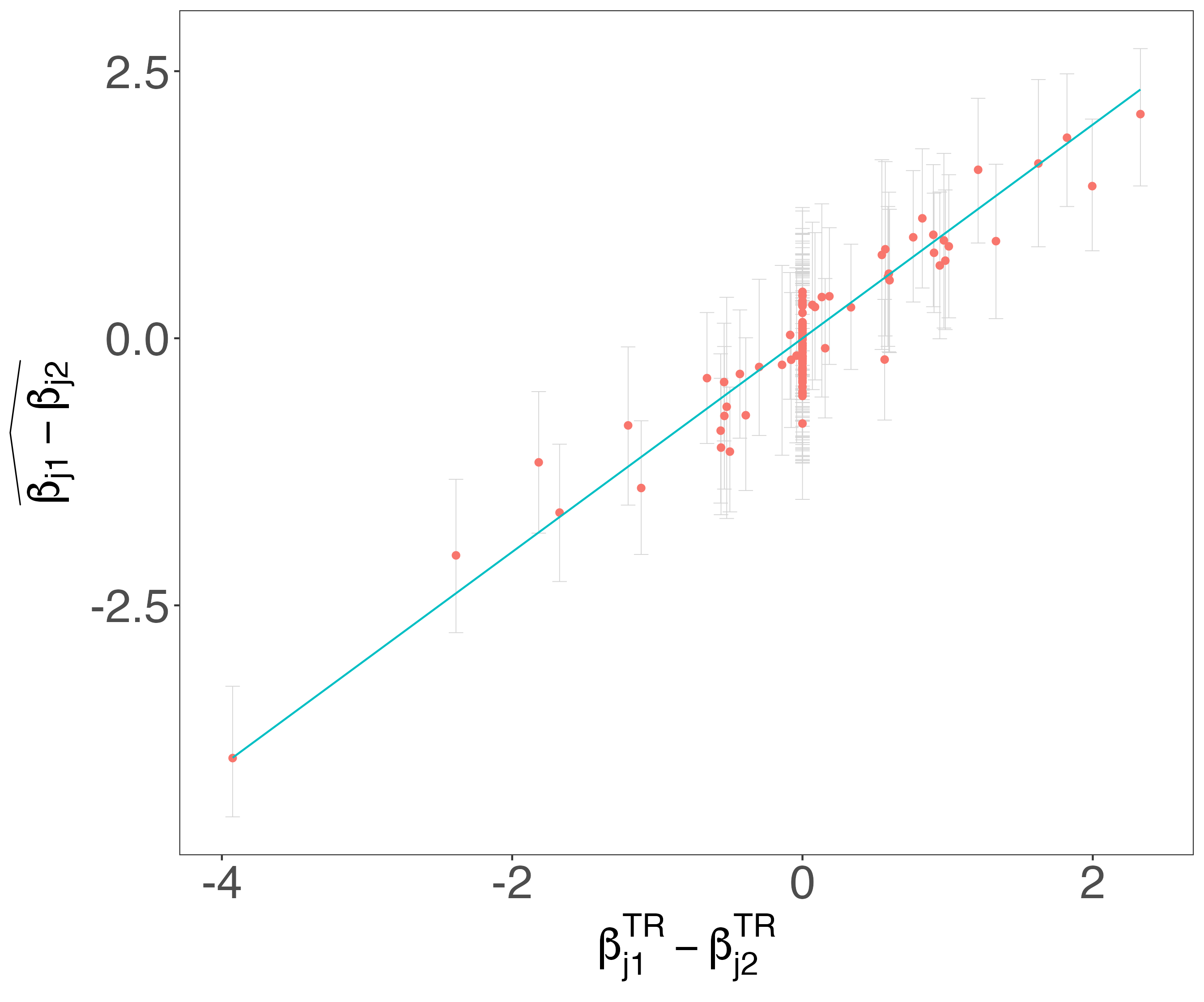}\\
   (a)  $\hat{\mu}_{ij}$ vs $\mu^\true_{ij}$ & (b) $\widehat{\beta_{j1}-\beta_{j2}}$ vs $\beta^\true_{j1}-\beta^\true_{j2}$ \\
  \end{tabular}
 \end{center}
 \vspace{-0.15in}
 \caption{[Simulation 3] Panels (a) and (b) compare the posterior estimates of $\mu_{ij}$ and $\beta_{j1} - \beta_{j2}$ to their true values. The dots represent the posterior median estimates, while the vertical lines indicate the 95\% credible interval estimates.
 } 
\label{fig:ch4-sim2-F2}
\end{figure}

Fig~\ref{fig:ch4-sim2-F1}(a) has a histogram of the differences $\hat{\Sigma}_{jj^\prime}(x^d) - \Sigma^\true_{jj^\prime}(x^d)$, $j \leq j^\prime$ for $x^d \in \{0, 1\}$. Fig~\ref{fig:ch4-sim2-F1}(b) and (c) compares the posterior estimates $\hat{\Sigma}(0)$ and  $\hat{\Sigma}(1)$ under our model to their truth. Arbitrary $\Sigma^\true(x^d)$ are generated for each condition, while our model has a covariance regression with low rank assumption. However, the model approximates the true interaction structure in $\Sigma^\true(x^d)$ reasonably well. Fig~\ref{fig:ch4-sim2-F2}(a) shows a posterior median estimate of mean abundance $\mu_{ij}$ with their 95\% interval estimate.  Panel (b) of the figure compares the posterior estimates of the covariate effect $\beta_{j1}-\beta_{j2}$ on the mean abundance to the truth.  The dots are point estimates, and the vertical lines 95\% credible interval estimates. Our model effectively captures the covariate effects on the mean abundance. 

To benchmark its performance, we compared it with MOFA+, SparCC, and CCLasso. For these comparators, we applied the methods separately for each covariate level, obtaining two covariance (or correlation) estimates per method. As shown in Table~\ref{tab:rmse-comparison}, our model achieves the smallest value of RMSE. While SparCC performed best among the comparators, it captured group-specific dependencies less accurately than our model (see Supp Fig 8).

\section{Mice Gut Microbiome Data Analysis}\label{sec:ch4-data}

We applied our method to a subset of the mice gut microbiome data from \cite{patnode2019interspecies}. The experiment aimed to understand how individual species in a human gut microbial community interact with others and respond to dietary changes. Specifically, it examined the effects of a human diet representing the upper tertile of saturated fat and lower tertile of fruit and vegetable consumption (HiSF) supplemented with different fibers. The abundance of beneficial gut microbes may change differently with fiber supplements, and their competition with other microbes may be influenced by the type of supplement.
In particular, gnotobiotic mice colonized with a 15-member consortium of human–gut–derived bacterial strains were fed the HiSF diet supplemented with different types of food-grade fibers. Each mouse received a different fiber-supplemented diet for a total of 4 weeks. The abundance of each community member was quantified using short-read shotgun DNA sequencing (community profiling by sequencing (COPRO-Seq)) \citep{hibberd2017intestinal, mcnulty2013effects} to determine the effects of a fiber treatment. 
For our analysis, we focused on the microbial composition at day 12 post-colonization, comparing three diets: two different fiber supplements, 10\% citrus pectin (CPT) and 10\% pea fiber (PEF), and HiSF as a control. This resulted in $N = 69$ samples, with 30, 20, and 19 mice assigned to the three diets: HiSF, CPT, and PEF, respectively. 
In addition, one specific species, OTU 2, corresponding to \emph{Bacteroides cellulosilyticus}, hereafter referred to as WH2), was omitted from half of the mice at the time of colonization to assess how its presence or absence affects microbial community dynamics.

\begin{figure}[t!]
  \begin{center}
\begin{tabular}{ccc}
    \includegraphics[width=.31\textwidth]{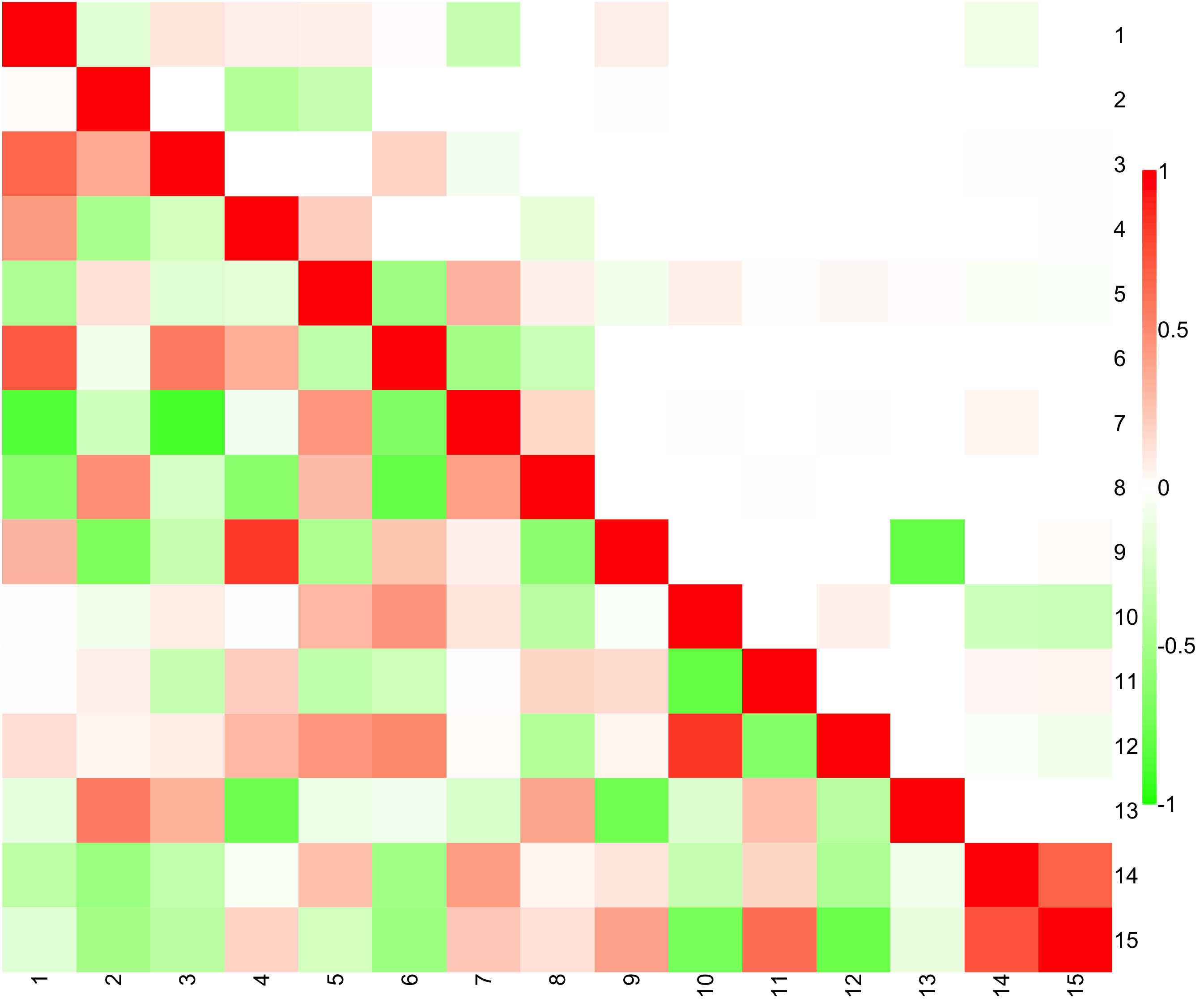} & 
   \includegraphics[width=.31\textwidth]{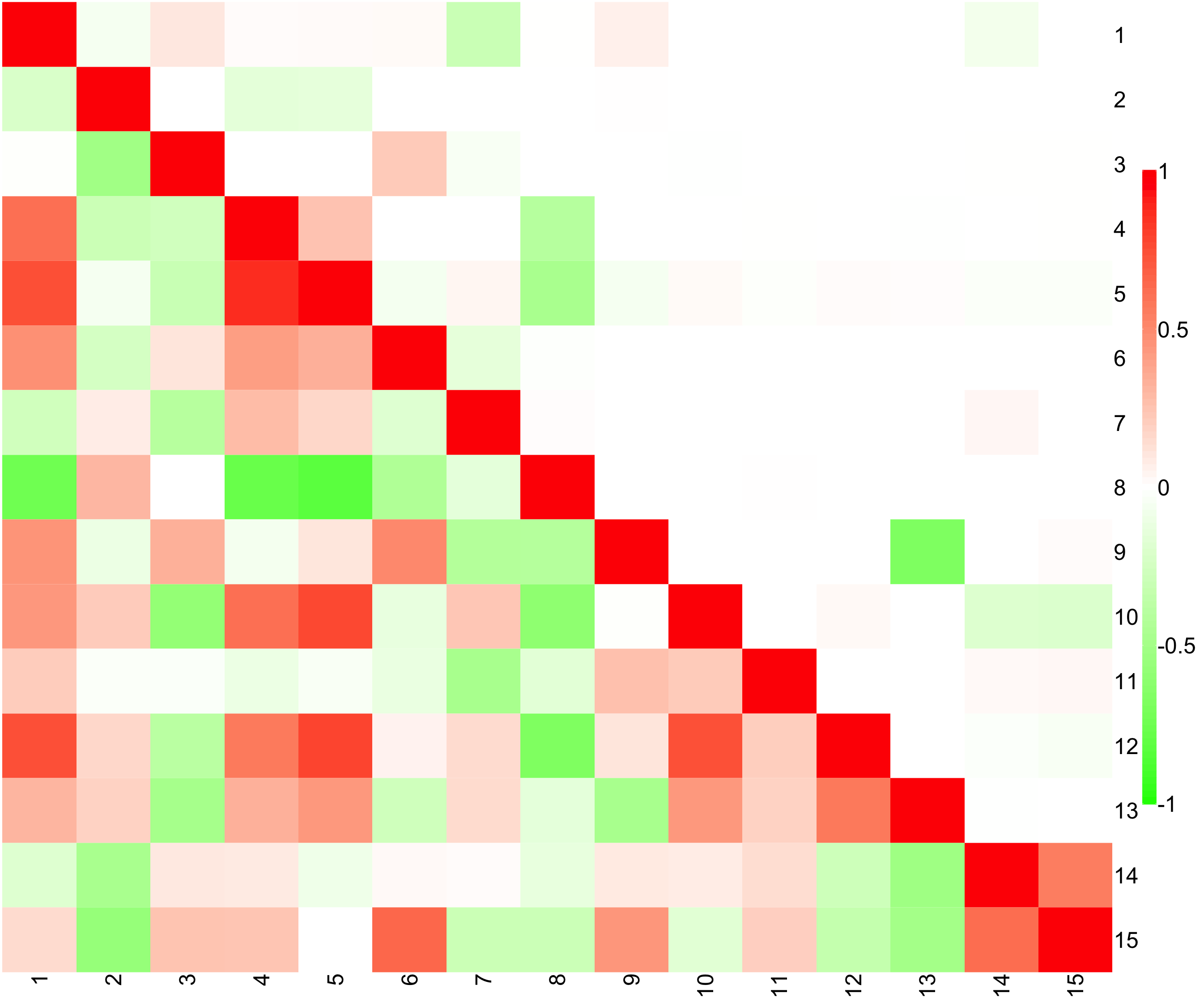} &
   \includegraphics[width=.31\textwidth]{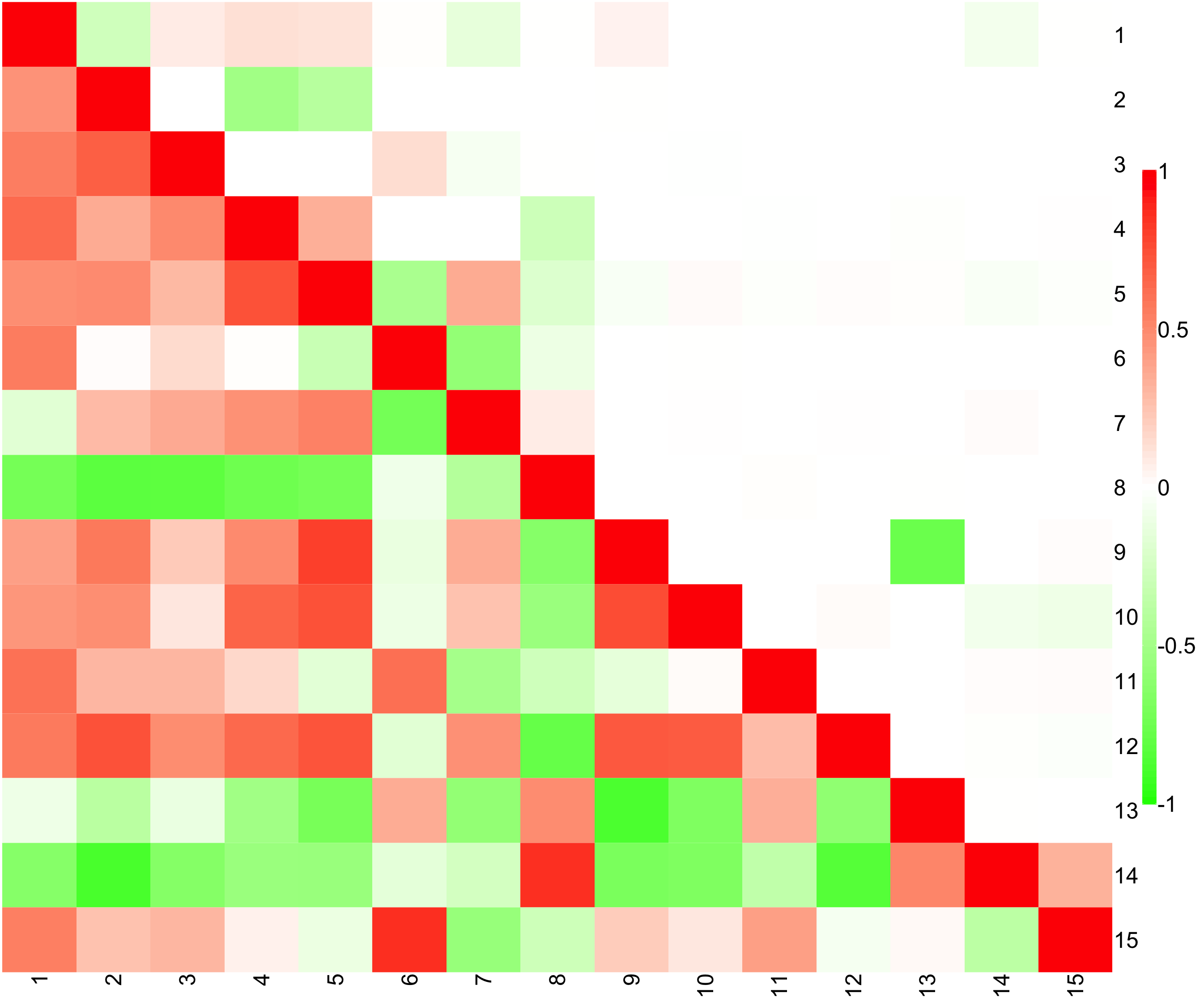} \\
   (a) absence of WH2 + CPT & (b) absence of WH2 + HiSF & (c) absence of WH2 + PEF\\
   \includegraphics[width=.31\textwidth]{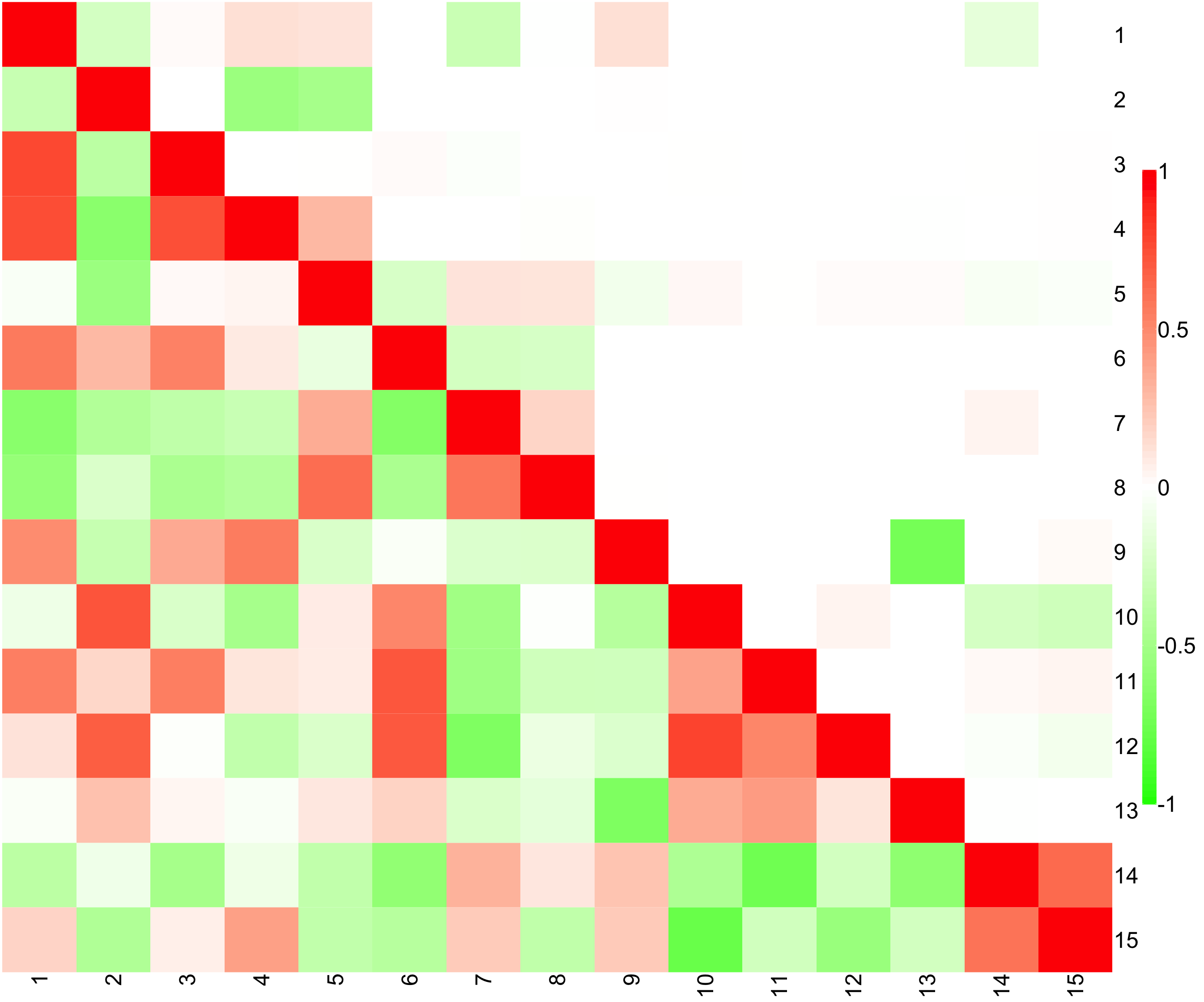} & 
   \includegraphics[width=.31\textwidth]{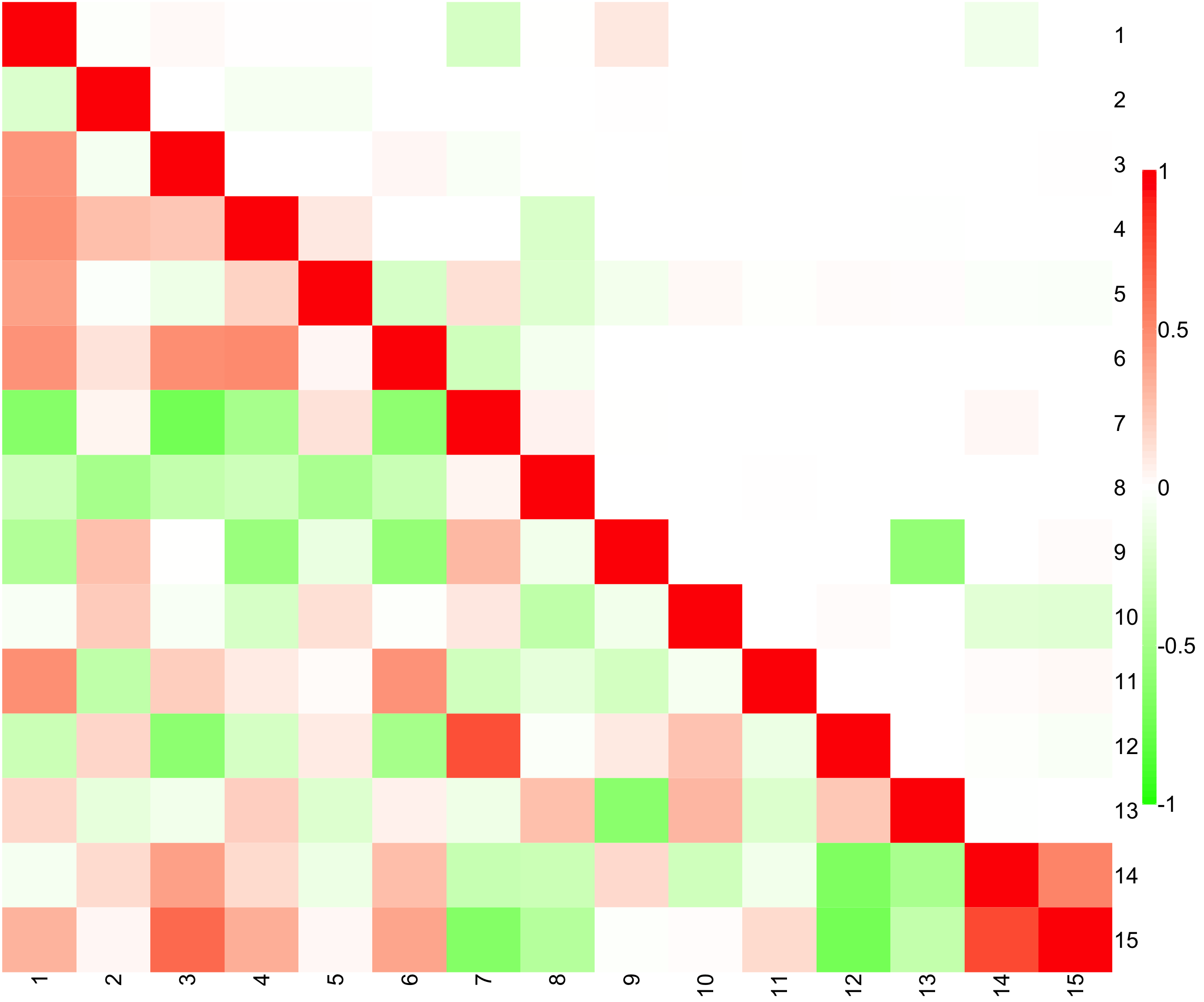} &
   \includegraphics[width=.31\textwidth]{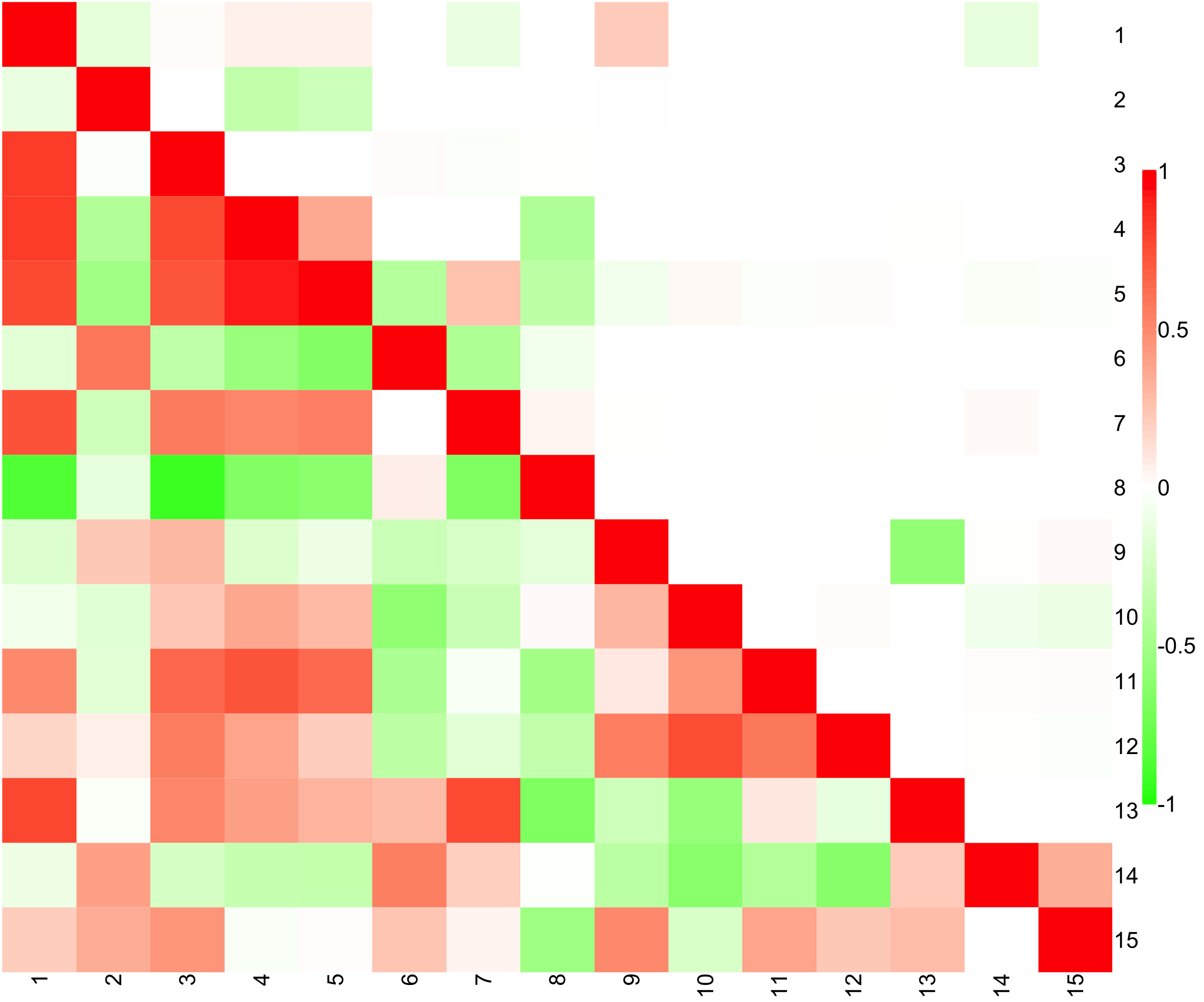} \\
   (d) CPT & (e) HiSF & (f) PEF\\
  \end{tabular}
 \end{center}
 \vspace{-0.15in}
 \caption{[Mice Data] The lower left and upper right triangles of the heatmap illustrate empirical correlation estimates $\rho^{\mbox{\footnotesize em}}_{jj^\prime}(\bx)$ and their posterior estimates $\hat{\rho}_{jj^\prime}(\bx)$ under the six different experimental conditions, respectively. } 
\label{fig:ch4-mice-corr}
\end{figure}

For analysis, we consider two categorical covariates for  $\mu$ and $\Sigma$: diet, which has three levels, and WH2 status, which is binary.  A list of the 15 strains in the defined bacterial consortium can be found in Supp.\ Tab.\ 2, and the number of samples for each of the six experimental conditions defined by diet and the WH2 status is provided in Supp.\ Tab.\ 3.  We note that even in samples where WH2 was omitted, the read counts for WH2 are non-zero due to background sequencing and mapping artifacts, similar to other OTUs. Therefore, we proceeded with all 15 OTUs for the analysis. 
%
The lower triangle of the heatmaps in Fig~\ref{fig:ch4-mice-corr} (a)-(f) shows empirical correlation estimates, $\rho^{\mbox{\footnotesize em}}_{jj^\prime}(\bx)$, computed using $\log(y_{ij} + 0.01)$ after normalization by the log of the total count sample size factor estimates for each of the six conditions. 
To fit our model, 
%
the fixed hyperparameters were set similarly to those in Simulation 1, with $K = 8$. The choice of $K$ can be determined by the PCA of transformed data as described in \S~\ref{sec:ch4-mcmc}. The MCMC simulation ran for 160,000 iterations, discarding the first half as burn-in and using the second half for inference. The computation took 22 minutes on an M1 Mac.

\begin{figure}[t!]
  \begin{center}
\begin{tabular}{cc}
   \includegraphics[width=.43\textwidth]{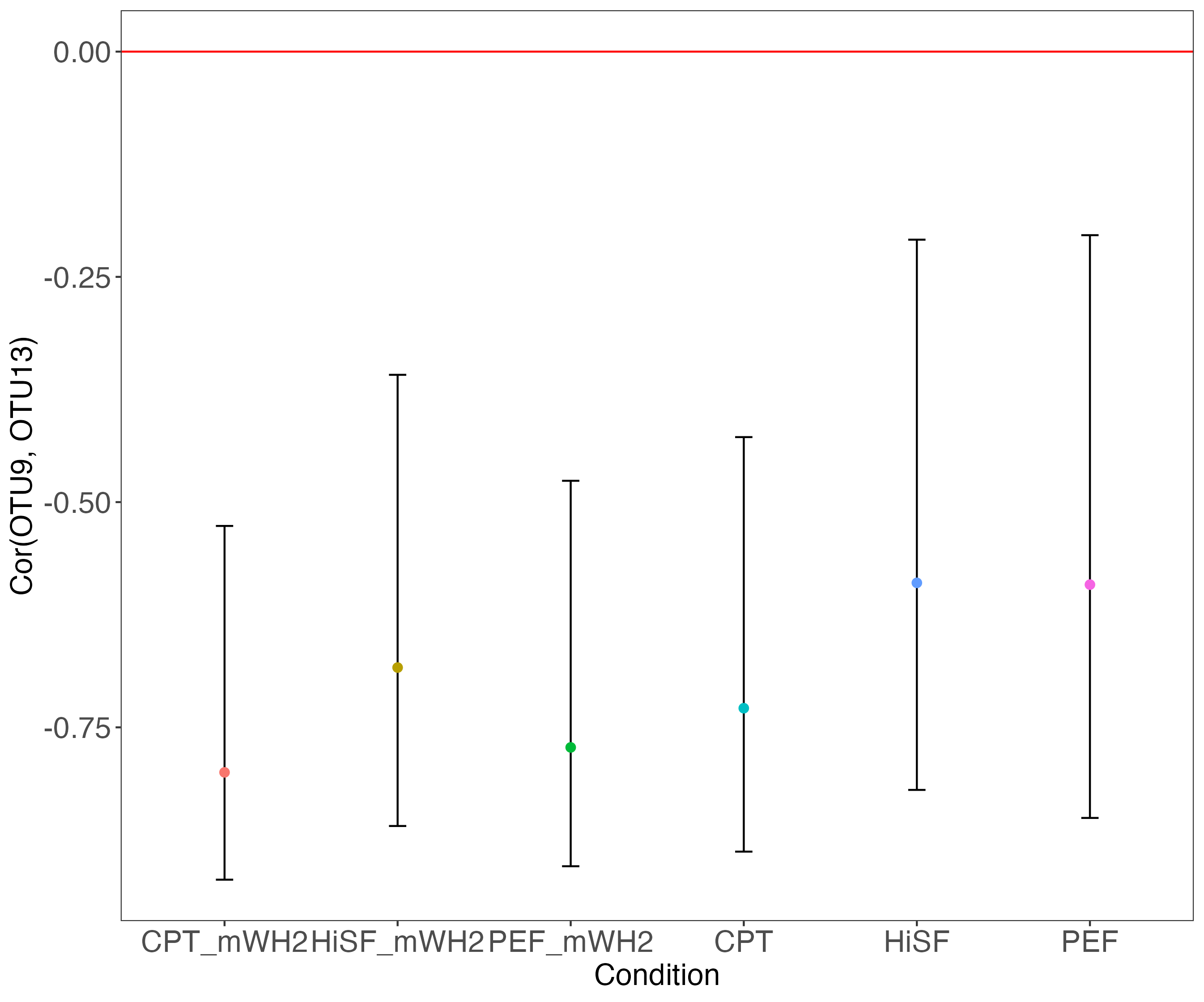} &
   \includegraphics[width=.43\textwidth]{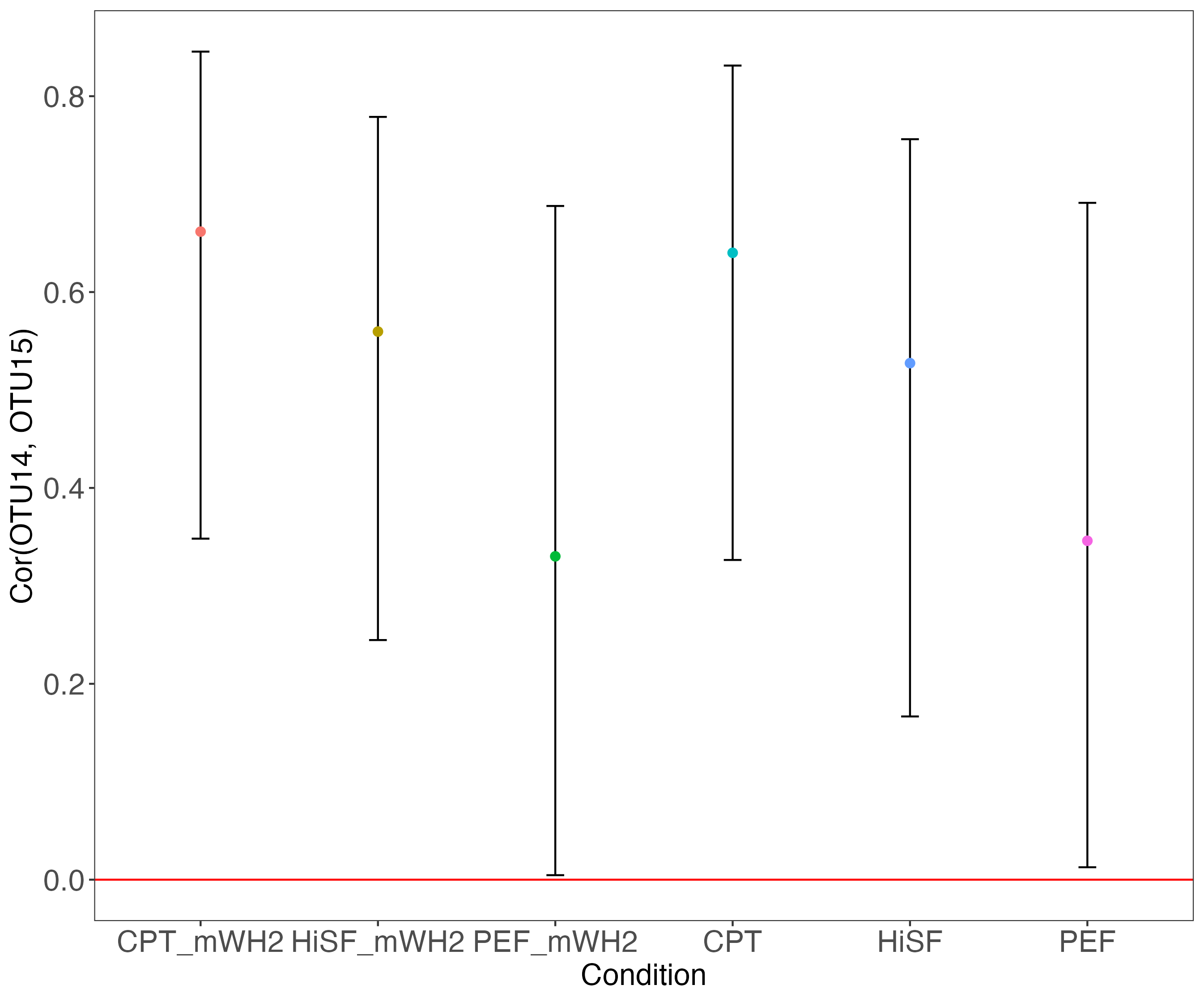} \\ 
   (a) OTUs 9 and 13 &
   (b) OTUs 14 and 15 \\
  \end{tabular}
 \end{center}
 \vspace{-0.15in}
 \caption{[Mice Data] Posterior inference on correlations $\rho_{jj^\prime}(\bx)$ under the six experimental conditions are illustrated for three pairs. The dots denote posterior median estimates, and the vertical lines 95\% credible interval estimates. } 
\label{fig:ch4-mice-corr-2}
\end{figure}

The upper triangle of the heatmaps in Fig~\ref{fig:ch4-mice-corr}(a)-(f) illustrates posterior mean estimates of correlation $\hat{\rho}_{jj^\prime}(\bx)$ of the OTUs under the six different conditions for easier interpretation. 
Overall, the interactions are sparse across all experimental conditions, and their structure varies with both diet and WH2 status. In particular, the diet modifications involve complex fiber preparations containing diverse polysaccharides; thus, they have the potential to disrupt interactions between species as they compete for distinct glycan structures present in PEF or CPT.
Fig~\ref{fig:ch4-mice-corr-2} illustrates posterior inference on $\rho_{jj^\prime}(\bx)$ for two selected pairs of species, where the dots and vertical lines represent the posterior median estimates and 95\% credible interval estimates, respectively.   
From panel (a), the 95\% posterior credible interval for $\rho_{jj'}(\mathbf{x})$ between OTUs 9 and 13, corresponding to \emph{Collinsella aerofaciens} and \emph{Ruminococcaceae} sp., indicates that their abundances are negatively associated under all conditions, with statistical significance.
Considering the large phylogenetic distance between these organisms (belonging to distinct phyla), the strong negative correlation between them across all treatment groups suggests that they may compete for a nutrient present in the background diet, such as protein breakdown products \citep{wolf2019bioremediation}.  
Panel (b) displays the results for the pair of OTUs 14 and 15 corresponding to \emph{Ruminococcus albus} and \emph{Subdoligranulum variabile}. The interaction between OTUs 14 and 15 is significantly positive across all conditions, as the 95\% credible intervals are strictly above zero. The estimated strength of their association varies across diets but shows little change with the status of WH2. 
%
%
%
%
Several additional interactions may indicate competition for polysaccharides, although these are not statistically significant. OTU 1 (\textit{B. ovatus}) and OTU 7 (\textit{B. finegoldii}), for which the results in Figure~\ref{fig:ch4-mice-corr} indicate a potentially negative association, likely compete for homogalacturonan, a carbohydrate found in the fruits and vegetables of the HiSF diet. A potential positive interaction between OTU 5 (\textit{B. vulgatus}) and OTU 7, especially in the CPT and PEF groups, is also notable. While these species can metabolize homogalacturonan, the interaction suggests that they might cooperatively break down this carbohydrate rather than compete for it.


\begin{figure}[!t]
  \begin{center}
\begin{tabular}{cc}
   \includegraphics[width=.45\textwidth]{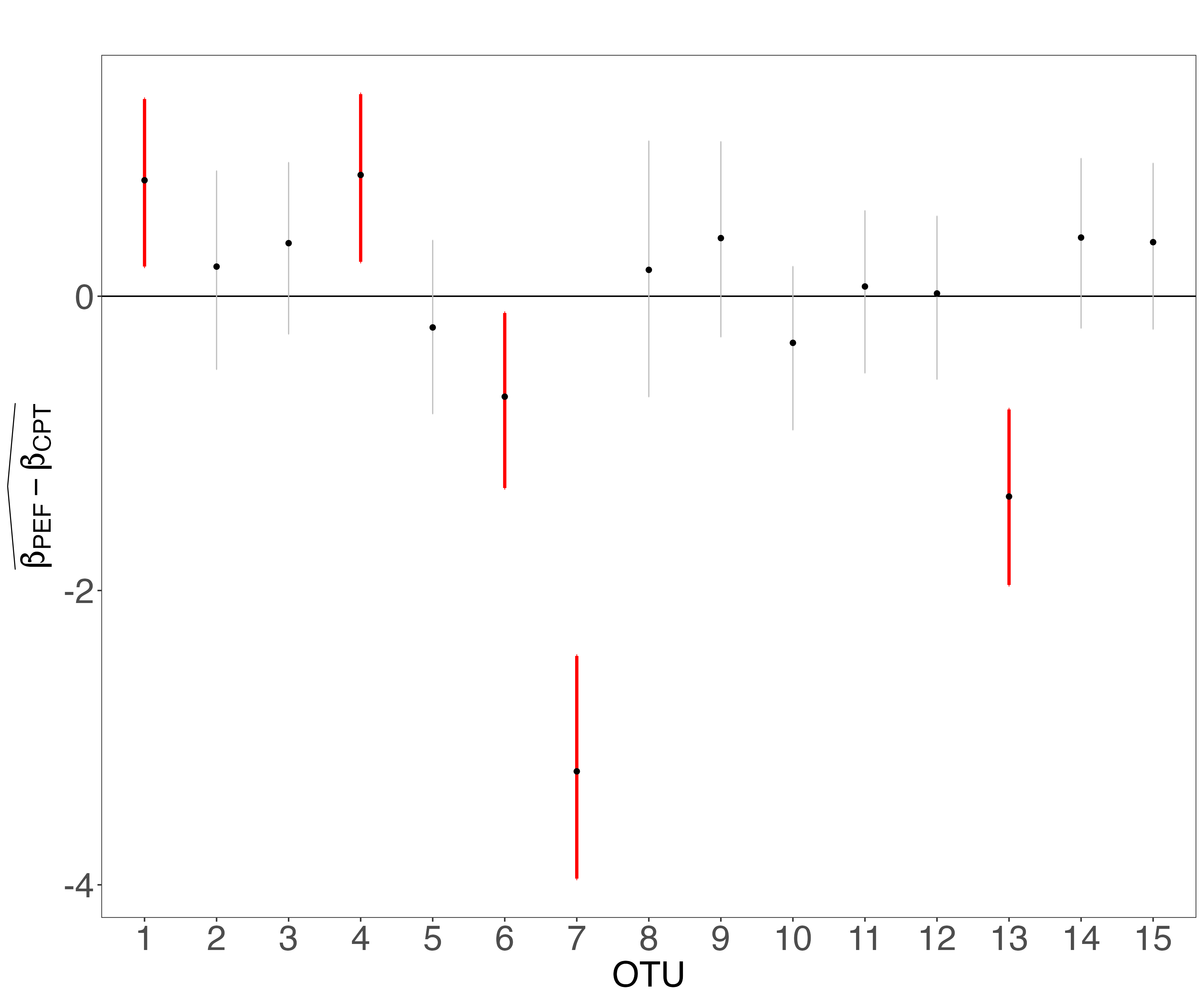} &
   \includegraphics[width=.45\textwidth]{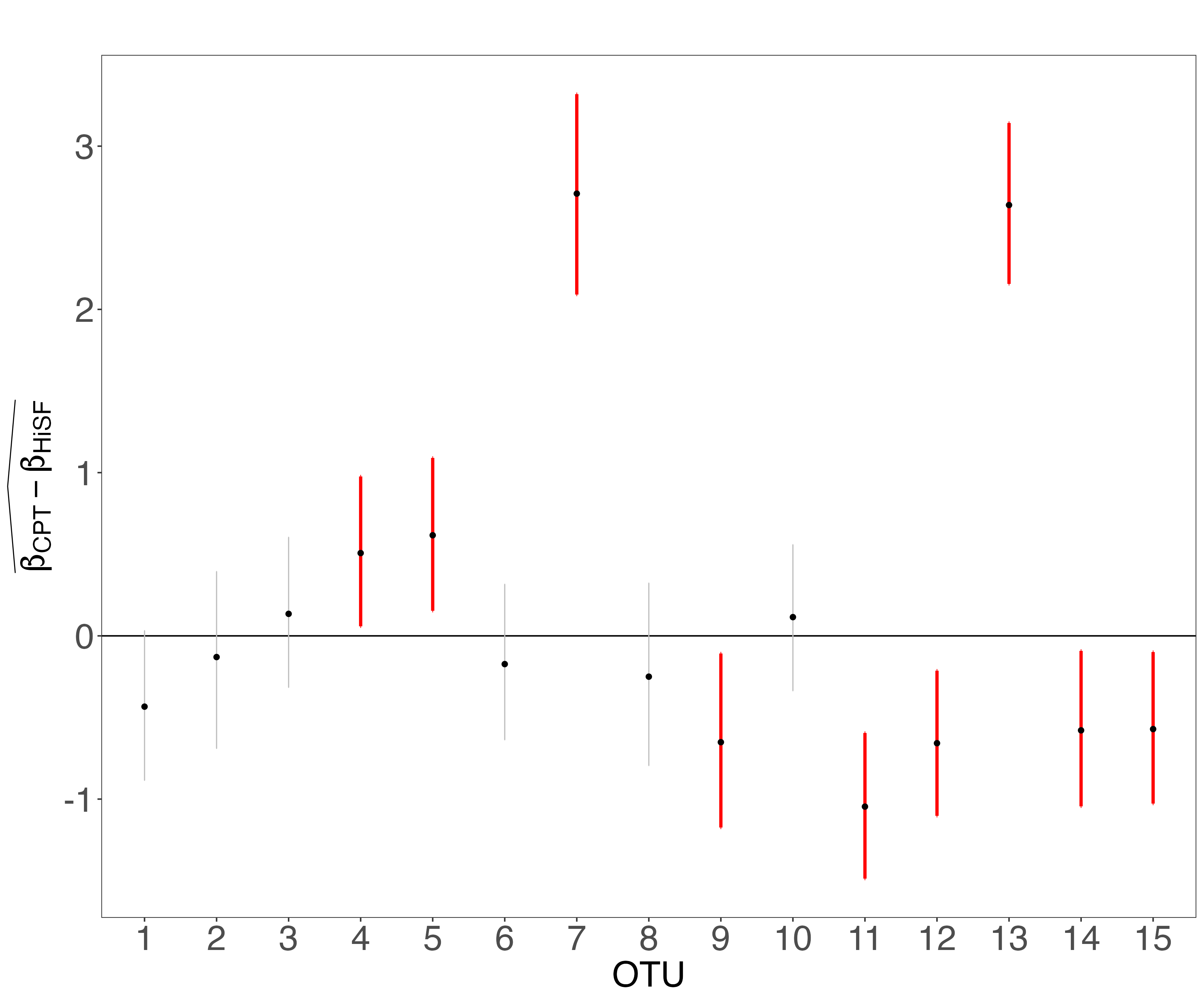}\\ (a) $\beta_{PEF}-\beta_{CPT}$ &
   (b)  $\beta_{CPT}-\beta_{HiSF}$\\
   \includegraphics[width=.45\textwidth]{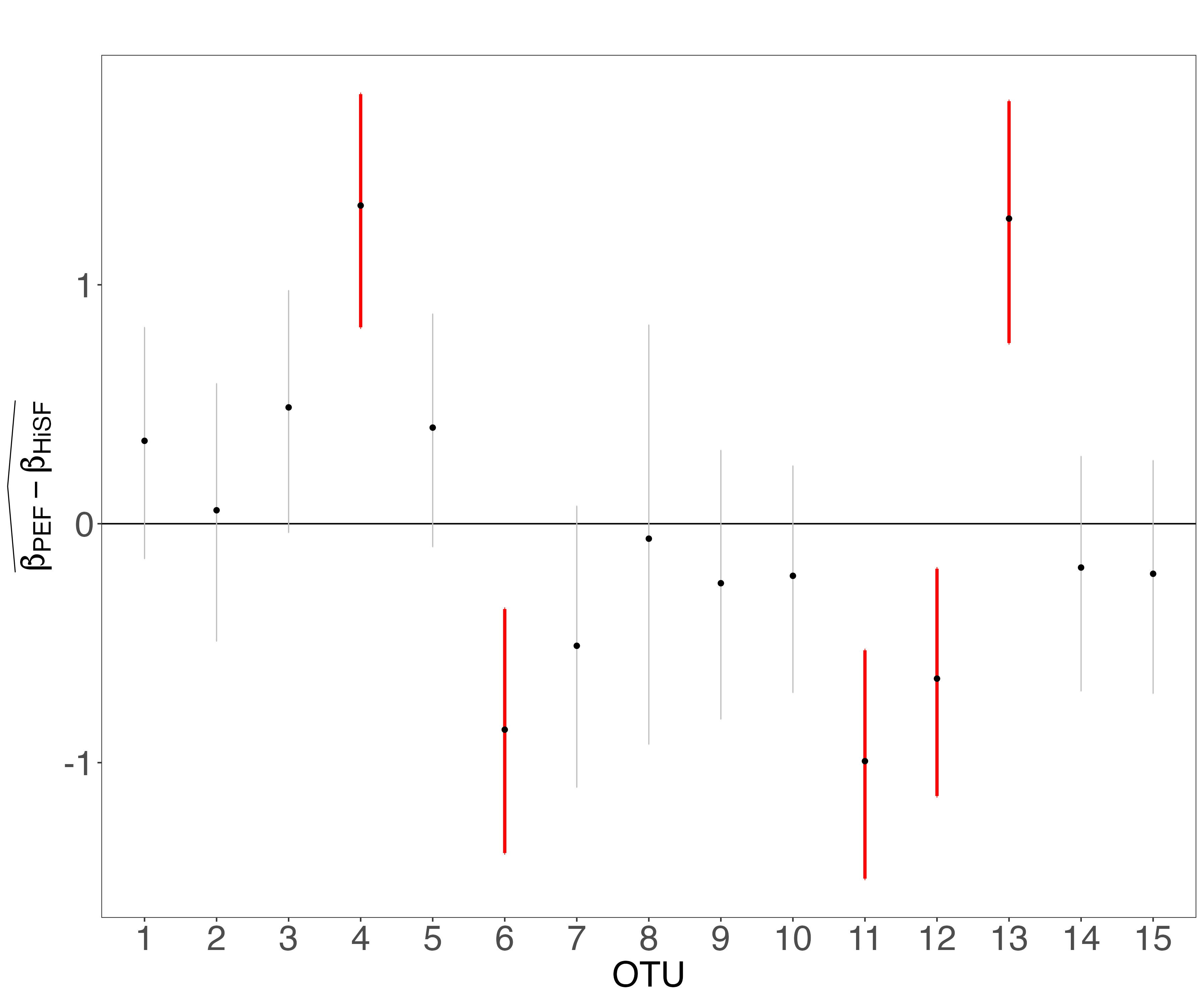}&
   \includegraphics[width=.45\textwidth]{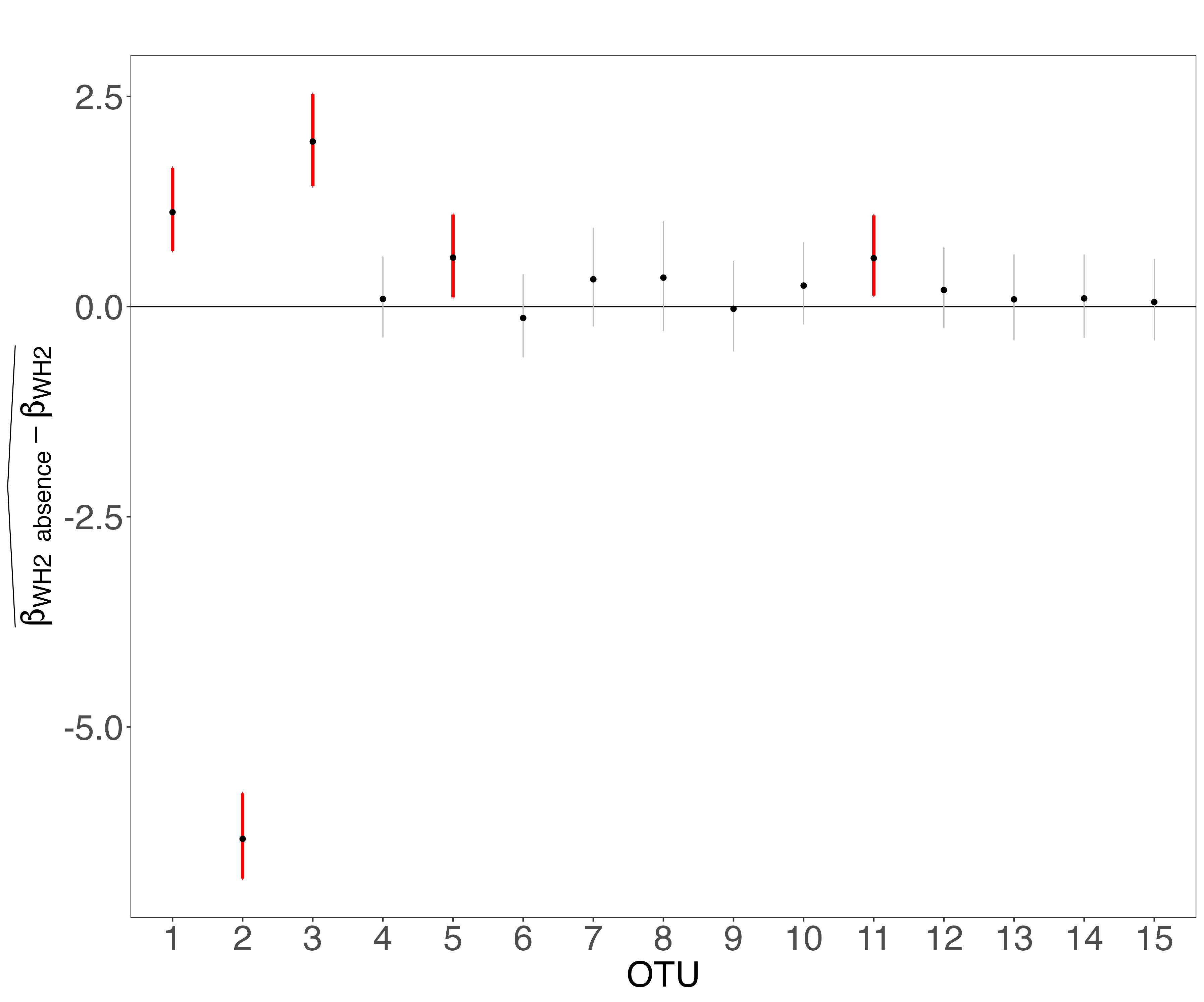}\\
   (c) $\beta_{PEF}-\beta_{HiSF}$ &
   (d) $\beta_{WH2~absence}-\beta_{WH2}$\\
  \end{tabular}
 \end{center}
 \vspace{-0.15in}
 \caption{[Mice Data] Posterior estimates of diet effects on mean abundances are illustrated.  Posterior median estimates are denoted by dots, and 95\% credible estimates with vertical lines. The intervals that do not contain zero are marked in red.} 
\label{fig:ch4-mice-beta}
\end{figure}

Fig~\ref{fig:ch4-mice-beta} illustrates posterior inferences on the differential microbial mean abundances across diets and WH2 status. The dots and vertical lines represent posterior median estimates and 95\% credible intervals, with intervals not containing zero highlighted in red. From panel (a), the diet supplemented with pea fiber is associated with statistically significant changes in abundance for five OTUs compared to the diet supplemented with citrus pectin. 
Panels (b) and (c) compare the control diet to the citrus pectin and pea fiber diets, respectively. From the figure, the abundance of \emph{B.\ thetaiotaomicron} (OTU 4), and \emph{Ruminococcaceae} (OTU 13) changes significantly across all pairwise diet comparisons.  Finally, panel (d) displays the impact of WH2 absence. The absence of WH2 is associated with a significant increase in the abundance of OTUs 1, 3, 5, and 11, corresponding to \emph{Bacteroides ovatus}, \emph{Bacteroides thetaiotaomicron} 7330, \emph{Bacteroides vulgatus}, and \emph{Odoribacter splanchnicus}, respectively. 
The initial analysis in \cite{patnode2019interspecies} also found significant differences in the relative abundance of OTU 4 across all three pairwise diet comparisons. These changes were mirrored by changes in protein expression, where OTU 4 demonstrated increased expression of enzymes for metabolizing pea fiber arabinan. 
OTUs 7 and 13 also showed a marked increase in relative abundance when CPT was added to the diet, confirming the results of the previous analysis. 
The finding of the expansion of OTUs 1 and 5 when WH2 is omitted from this community is notable as it was  observed in multiple independent gnotobiotic mouse experiments.

We further analyzed the real data using MOFA+, sparCC, and CCLasso, with the results presented in Supp.\ Figs.~9–11. MOFA+ was implemented with covariates included, as it accommodates categorical variables. In contrast, sparCC and CCLasso were applied separately within each group because they do not accommodate covariates. Note that these methods provide inference only on interactions.

\section{Conclusion}\label{sec:ch4-con}

In this paper, we developed a Bayesian joint model of mean and covariance varying with covariates for high-dimensional multivariate count data. This method utilizes a covariate-varying factor model for the covariance matrix and also models the mean abundance using a flexible DP mixture. The method thus enables the assessment of covariate effects on mean and covariance in tandem. We placed a Dirichlet-Horseshoe prior on the covariate-varying loading matrix to induce sparse feature interactions. The flexible mean mixture kernels handle the excess zeros and over-dispersion problems in the count data. Its performance is demonstrated through simulations and a real data example. From a methodological perspective, our model contributes to the growing literature on covariance regression by extending it to high-dimensional count settings and embedding it within a nonparametric Bayesian framework. Unlike existing approaches that treat covariance as constant over covariates or rely on Gaussian models on transformed data, our method captures dynamic dependence patterns directly in discrete data without ad hoc transformations.

Our methods can be further extended by relaxing the linear covariance regression to a more complicated regression formula, such as introducing the transformation of continuous covariates $\sqrt{x}$ and $x^\frac{3}{2}$. Introducing different orders of covariates induces a higher order of covariance-covariate relationship. It is important to note that while this extension offers greater flexibility, the same higher order of covariates entering the mean regression would need more exploration, e.g., variable selection or shrinkage estimators. With added parameters, more samples are needed to obtain reasonable inferences. A further area of research is to study interactions that vary over time and/or space using temporal or spatial multivariate count data. In the presence of temporal or spatial heteroscedasticity, it is natural to consider conditional dependence across time points or locations. \cite{fieuws2006pairwise} discussed a pairwise approach for jointly modeling multivariate longitudinal data using mixed models, providing a foundation for understanding covariance structures in such settings. Adapting the factor loading matrix to be time- or location-dependent has the potential to enhance inference on interaction structures in other domains.

\bibliographystyle{apalike}

\bibliography{Reference}

\end{document}